\documentclass[a4paper,11pt]{article}
\usepackage{jheppub} 
\usepackage{orcidlink}

\usepackage{subfigure}
\usepackage{mathtools}
\usepackage{amsmath}
\usepackage{amsfonts}
\usepackage{amssymb}
\usepackage{bbold}
\usepackage{bm}
\usepackage{ esint }
\usepackage{orcidlink}

\newcommand{\br}{{\bf r}}

\newcommand{\bk}{{\bf k}}

\newcommand{\bv}{{\bf v}}

\newcommand{\bp}{{\bf p}}

\newcommand{\tomegaE}{{\tilde{\omega}_{E}}}
\newcommand{\omegaE}{{\omega_{E}}}

\newcommand{\sA}{{\sf A}}
\newcommand{\sC}{{\sf C}}

\newcommand{\sH}{{\sf H}}
\newcommand{\sM}{{\sf M}}
\newcommand{\sN}{{\sf N}}

\newcommand{\oP}{{\overline{P}}}

\newcommand{\opsi}{{\overline{\psi}}}

\newcommand{\bpsi}{{\boldsymbol{\psi}}}

\newcommand{\bpartial}{\boldsymbol{\partial}}

\newcommand{\cV}{c_{\rm V}}
\newcommand{\sV}{s_{\rm V}}
\newcommand{\tV}{\theta_{\rm V}}
\newcommand{\chiI}{\chi_{\rm I}}

\newcommand{\exclude}[1]{{}}
\long\def\exclude#1{}

\newcommand{\GF}{G_{\rm F}}

\title{Theory of neutrino slow flavor evolution.
Part I. Homogeneous medium.}

\author[a]{Damiano F.\ G.\ Fiorillo \orcidlink{0000-0003-4927-9850}} 
\affiliation[a]{Deutsches Elektronen-Synchrotron DESY,
Platanenallee 6, 15738 Zeuthen, Germany}

\author[b]{and Georg G.\ Raffelt
\orcidlink{0000-0002-0199-9560}}
\affiliation[b]{Max-Planck-Institut f\"ur Physik, Boltzmannstr.~8, 85748 Garching, Germany}

\abstract{Dense neutrino gases can exhibit collective flavor instabilities, triggering large flavor conversions that are driven primarily by neutrino-neutrino refraction. One broadly distinguishes between fast instabilities that exist in the limit of vanishing neutrino masses, and slow ones, that require neutrino mass splittings. In a related series of papers, we have shown that fast instabilities result from the resonant growth of flavor waves, in the same way as turbulent electric fields in an unstable plasma. Here we extend this framework to slow instabilities, focusing on the simplest case of an infinitely homogeneous medium with axisymmetric neutrino distribution. The relevant length and time scales are defined by three parameters: the vacuum oscillation frequency $\omegaE=\delta m^2/2E$, the scale of neutrino-neutrino refraction energy $\mu=\sqrt{2}\GF(n_\nu+n_{\overline\nu})$, and the ratio between lepton and particle number $\epsilon=(n_\nu-n_{\overline\nu})/(n_\nu+n_{\overline\nu})$. We distinguish between two very different regimes: (i)~For $\omegaE\ll \mu \epsilon^2$, instabilities occur at small spatial scales of order $(\mu\epsilon)^{-1}$ with a time scale of order $\epsilon \omegaE^{-1}$. This novel branch of slow instability arises from resonant interactions with neutrinos moving along the axis of symmetry. (ii)~For $\mu \epsilon^2\ll \omegaE\ll \mu$, the instability is strongly non-resonant, with typical time and length scales of order $1/\sqrt{\omegaE \mu}$. Unstable modes interact with all neutrino directions at once, recovering the characteristic scaling of the traditional studies of slow instabilities. In the inner regions of supernovae and neutron-star mergers, the first regime may be more likely to appear, meaning that slow instabilities in this region may have an entirely different character than usually envisaged.}

\begin{document}
\maketitle
\flushbottom

\section{Introduction}\label{sec:introduction}

In laboratory settings, the flavor evolution of neutrinos consists of the usual oscillations caused by masses and mixing \cite{deSalas:2020pgw, Capozzi:2021fjo, Esteban:2020cvm}, although matter refraction can play an important role \cite{Wolfenstein:1977ue, Wolfenstein:1979ni}, notably for solar or supernova neutrinos that escape through a density gradient, engendering MSW conversion \cite{Mikheyev:1985zog, Mikheev:1986wj, Mikheev:1986if, Dighe:1999bi}. On the other hand, in neutrino-dense astrophysical environments, neutrino-neutrino refraction \cite{Pantaleone:1992eq} spawns very different modes of flavor conversion in the form of collective flavor waves supported by the interacting neutrino gas. The key insight was that these waves can be unstable and thus can lead to large degrees of self-induced flavor coherence \cite{Samuel:1993uw, Samuel:1995ri, Duan:2005cp, Duan:2006an} even without neutrino masses or mixing  \cite{Sawyer:2004ai, Sawyer:2008zs, Chakraborty:2016lct} except to seed the instabilities. Understanding these phenomena and their astrophysical relevance has remained an unfinished effort since their discovery some thirty years ago.

However, the underlying mean-field equations have remained the same. The flavor structure of the neutrino gas is represented by $3{\times}3$ flavor density matrices $\varrho(\bp,\br,t)$, where the diagonal elements are the usual occupation numbers $f_\alpha(\bp,\br,t)$ for flavor $\alpha$, whereas the off-diagonal elements, that we will denote as $\psi_{\alpha\beta}(\bp,\br,t)$, represent the amount of coherence between flavors $\alpha$ and $\beta$. (Here and always we work in the weak-interaction basis, not the mass basis.) The evolution is governed by the quantum-kinetic equation \cite{Dolgov:1980cq, Rudsky, Sigl:1993ctk, Sirera:1998ia, Yamada:2000za, Vlasenko:2013fja, Volpe:2013uxl, Serreau:2014cfa, Kartavtsev:2015eva, Fiorillo:2024fnl, Fiorillo:2024wej}
\begin{equation}\label{eq:QKE}
    (\partial_t+\bv\cdot\bpartial_\br)\varrho(\bp,\br,t)
=-i\bigl[\sH(\bp,\br,t),\varrho(\bp,\br,t)\bigr]
+{\sC}(\varrho,\overline\varrho)
\end{equation}
and a similar equation for the antineutrino modes $\overline\varrho(\bp,\br,t)$. The Liouville operator on the left-hand side, also denoted as Vlasov operator in the context of plasma physics, takes care of free streaming and involves the velocity which, in the ultrarelativistic limit, is $\bv=\bp/|\bp|$ and thus simply the direction of motion. The collision term ${\sC}(\varrho,\overline\varrho)$ depends on complicated convolutions of $\varrho(\bp,\br,t)$ and $\overline\varrho(\bp,\br,t)$ and the constituents of the nuclear medium. The refractive Hamiltonian, really a matrix of oscillation frequencies, is
\begin{equation}\label{eq:Hamiltonian}
    \sH(\bp,\br,t)=\pm\frac{\sM^2}{2E}+\sqrt{2}\GF\sN
    +\sqrt{2}\GF\int\frac{d^3\bp'}{(2\pi)^3}\,\bigl[\varrho(\bp',\br,t)-\overline\varrho(\bp',\br,t)\bigr](1-\bv'\cdot\bv),
\end{equation}
where $\sM$ is the neutrino mass matrix and the negative sign applies to antineutrinos. In the rest frame of the medium, matter refraction is determined by the matrix $\sN$ of net charged fermion densities, i.e., it has $n_{e^-}-n_{e^+}$ etc.\ on the diagonal. It is the commutator term that takes care of flavor mixing and neutrino-matter and neutrino-neutrino refraction. 

A brute-force numerical solution of Eq.~\eqref{eq:QKE} is usually out of the question. On the other hand, hierarchies of scales can simplify the problem. MSW conversion in the free-streaming regime is driven by a gradient of the matter density and the conversion is adiabatic when the vacuum oscillation frequency $\omega_E=\delta m^2/2E$ is fast by comparison. In the early days of collective oscillation studies in SNe
\cite{Duan:2009cd, Duan:2010bg}, a similar approach consisted of assuming a stationary emitting surface (the bulb model) and one looked for static solutions, i.e., time-independent variations of the neutrino flavor field as a function of radius. The gradient of the neutrino density then drove a nearly adiabatic evolution along the radius, spawning intriguing signatures such as spectral splits (or swaps).

It has long since emerged that a static slow variation along the radial direction was largely an artifact of too many symmetry assumptions. Unstable collective modes depend both on their spatial variation, possibly on very small scales relative to overall geometric ones \cite{Mangano:2014zda, Duan:2014gfa, Abbar:2015mca}, as well as their time variation \cite{Abbar:2015fwa, Dasgupta:2015iia, Capozzi:2016oyk, Mirizzi:2015fva}. A small-scale wave can be unstable in a neutrino gas where a large-scale one is stable. Therefore, a completely opposite philosophy has recently taken root, looking at self-induced flavor conversion as a local phenomenon relative to overall geometric scales, potentially as a basis for numerical implementation on subgrid scales~\cite{Bhattacharyya:2020jpj, Zaizen:2022cik, Nagakura:2023jfi, Xiong:2023vcm, Shalgar:2022lvv, Cornelius:2023eop, Abbar:2024ynh, Richers:2024zit}. So what are the relevant spatial and temporal scales?

The answer to this question has also evolved. The original run-away effect of flavor coherence, Samuel's bimodal oscillations \cite{Samuel:1995ri}, came from a feedback loop between vacuum oscillations and neutrino-neutrino refraction. A homogeneous and isotropic gas of $\nu_e\overline\nu_e$ with monochromatic energy $E$ periodically oscillates as $\nu_e\overline\nu_e\leftrightarrow\nu_x\overline\nu_x$ in a fashion similar to a pendulum with natural frequency $\sqrt{\omega_E\mu}$ \cite{Hannestad:2006nj}, where $\mu=\sqrt{2} \GF(n_{\nu}+n_{\overline\nu})$ is a measure for the neutrino-neutrino refractive effect. In contrast to the originally studied environment around cosmological neutrino decoupling, in the region of spatial neutrino decoupling near a SN core, the neutrino-antineutrino asymmetry is typically not small so that a new parameter appears that we express as $\epsilon=(n_{\nu}-n_{\overline\nu})/(n_{\nu}+n_{\overline\nu})$ and concomitant refractive energy shift $\epsilon\mu=\sqrt{2}\GF(n_\nu-n_{\overline\nu})$. In an isotropic monochromatic neutrino gas, the bimodal instability appears only for $\epsilon^2\mu<\omega_E<\mu$ and, unless $\epsilon$ is very small, $\omega_E$, $\mu$, and the growth rate $\sqrt{\omega_E\mu}$ are all of a similar general order. We are assuming here that the gas possesses initially mostly $\nu_e$ and $\overline{\nu}_e$; otherwise, these definitions should be changed as $n_{\nu}\to n_{\nu_e}-n_{\nu_x}$ and $n_{\overline{\nu}}\to n_{\overline{\nu}_e}-n_{\overline{\nu}_x}$.

Unstable modes also exist in the absence of neutrino masses. In this limit, the equation for lepton number (neutrinos minus antineutrinos) becomes self-contained and is a phenomenon that only involves the flavor field of lepton number, not particle number. This multi-angle effect was discovered in homogeneous systems consisting of a few discrete neutrino directions \cite{Sawyer:2004ai, Sawyer:2008zs} and generally requires a crossing of the angular flavor lepton number distribution \cite{Morinaga:2021vmc, Dasgupta:2021gfs, Fiorillo:2024bzm}. While historically the bimodal instability was discovered first, the multi-angle instability \cite{Sawyer:2004ai, Sawyer:2008zs, Sawyer:2015dsa, Chakraborty:2016lct, Izaguirre:2016gsx, Airen:2018nvp, Johns:2019izj, Padilla-Gay:2021haz,  Fiorillo:2023mze, Fiorillo:2023hlk, Fiorillo:2024qbl, Fiorillo:2024bzm, Fiorillo:2024uki} is actually more fundamental as it does not require neutrino masses and thus is the purest form of collective flavor evolution. The generic scale is $\epsilon\mu$, although for the required difference of neutrino and antineutrino distributions, a single parameter $\epsilon$ can only be taken as an approximate overall measure. 

A first taxonomy of different scales of flavor conversion was provided by Sawyer \cite{Sawyer:2008zs} who denoted bimodal oscillations as fast (scale $\sqrt{\omega_E\mu}$) relative to vacuum oscillations (scale~$\omega_E$), and the multi-angle effect as very fast (scale $\mu$), whereas the collision rate is the slowest of all scales. Today, motivated by the hierarchy $\sqrt{\omega_E\mu}\ll\mu$, bimodal conversion is termed slow flavor conversion, the multi-angle effect as fast flavor conversion (FFC), meaning the limit of vanishing neutrino masses. We stick to this terminology, although we have already argued that the actually relevant scales for the different effects in the same neutrino gas are much more complicated. Later, we will argue that the comparison with $\sqrt{\omega_E\mu}$ is not consistent to begin with, highlighting the subtleties in these distinctions based on the growth rate. Moreover, one needs to discriminate more carefully between spatial and temporal scales. Instabilities driven by both $\omega_E$ and angular crossings have only recently been considered at all \cite{Airen:2018nvp, DedinNeto:2023ykt}. For us, slow instabilities truly mean the ones that become stable when $\omega_E$ is ``adiabatically'' driven to zero.

Another scale is matter refraction expressed by $\lambda=\sqrt{2}\GF(n_{e^-}-n_{e^+})$ if only charged leptons of the electron flavor are present. This scale is usually much larger than all others, exceeding $\omega_E$ by some 11 orders of magnitude in a SN core and implying that neutrino eigenstates of propagation and those of interaction nearly coincide. Therefore, the amplitude of mass-driven oscillations are strongly suppressed as first emphasized by Wolfenstein~\cite{Wolfenstein:1979ni}, justifying the traditional neglect of flavor conversion in SN simulations. On the other hand, the matter effect does not suppress collective flavor instabilities, essentially leading to a common rotation of all modes in flavor space \cite{Duan:2006an}. Still, one consequence is that neutrino masses drive instabilities not by directly inducing conversions, but rather by causing an energy splitting between neutrinos and antineutrinos, and therefore only the mass term projected on the flavor axis $\tomegaE=\omega_E\cos2\tV$, where $\tV$ is the vacuum mixing angle, actually causes the dynamics. The matter effect also modifies how unstable modes are triggered by the mass term, potentially by matter inhomogeneities that communicate seeds to the neutrinos on different length scales \cite{Airen:2018nvp}. Even neutrinos alone presumably contain inhomogeneities; the term that sources the instability is proportional to the neutrino density, which probably has small fluctuations on short length scales that would seed instabilities on these scales. However, in most studies, matter refraction and vacuum mixing are both ignored and instead an arbitrary seeding of instabilities is introduced. A first explicit study of the matter effect for FFC is not entirely conclusive~\cite{Sigl:2021tmj}.

Looking at collective flavor evolution as waves supported by the underlying neutrino directional and energy distribution \cite{Izaguirre:2016gsx, Capozzi:2017gqd, Yi:2019hrp} opens new perspectives \cite{Fiorillo:2024qbl, Fiorillo:2024bzm, Fiorillo:2024uki}. Strong analogies can be drawn from plasma physics where similar questions have come up and sometimes took decades to sort out. If we consider a flavor wave characterized by a real wave number $\bk$ and a potentially complex frequency $\omega$, we can define the complex phase velocity $u=\omega/|\bk|$, where homogeneous modes ($|\bk|=0$) have infinite and therefore strongly superluminal $u$. On the other hand, subluminal modes have the crucial property that they are on resonance with those neutrinos that have velocities $\bv$ such that they are on resonance with the wave. This Cherenkov condition allows for the exchange of energy between individual neutrinos and the collective wave and can lead either to Landau damping or exponential growth. Physical subluminal waves are therefore either Landau damped or grow exponentially, these being alternatives, not complex-conjugate solutions with a growing and damped branch. For superluminal waves, on the other hand, the Cherenkov condition cannot be fulfilled and one has either real $\omega$ or two complex conjugate solutions. The resonance picture is at the core of our proof that an angular crossing guarantees fast flavor instabilities of modes with $\bk$ pointing in the direction of a crossing line \cite{Fiorillo:2024bzm}. 

Our new perspective on flavor waves, in the linear regime, as analogous to plasma waves and concomitant Cherenkov-type energy transfer between collective waves and particles, was developed for fast flavor waves \cite{Fiorillo:2024bzm, Fiorillo:2024uki} and will be here extended to slow modes. In other words, we are seeking the dispersion relation for spatial Fourier modes with real $\bk$ and concomitant real or complex frequency $\omega$. This approach delivers, in the linear regime, the behavior of flavor waves in an infinite medium that we will take to be initially homogeneous. In principle, one could also include a spectrum of spatial inhomogeneities in the matter and/or the neutrino density, but not global gradients. The scales of the modes entering the problem are not a priori obvious because all of $\omega_E$, $\mu$, $\epsilon\mu$ and $\epsilon^2\mu$ could enter instead of a single scale $\epsilon\mu$ in fast flavor physics. On the other hand, even the smallest of these scales in the form of $\omega_E$ corresponds to an inverse length scale of kilometers (based on the atmospheric mass splitting) and thus remains somewhat small relative to geometric SN~scales, so the assumption of modes spatially small compared to global SN scales is generally justified. In addition, to cleanly separate slow instabilities from fast instabilities, we restrict ourselves here to situations where the energy-integrated lepton number does not have angular crossings, i.e., it does not change sign across different directions. In this way, we ensure that for $\omega_E\to 0$ there are no fast instabilities. We leave for future work the interplay between fast and slow instabilities in the case of a crossed angular distribution. 

What is the physical relevance of such studies? We are looking at the linear regime and concomitant scales of a time-dependent problem in infinite space, which however is meant to represent a small volume on SN scales. In the nonlinear regime, this picture corresponds to the subgrid volumes that have been numerically studied to understand the possible local relaxation of the neutrino flavor field within the limits of flavor lepton conservation. The main assumption in this approach is that the unstable flavor waves grow nonlinear and possibly relax to equilibrium in the same small volume in which they were born.

As stressed earlier, this approach is entirely opposite to early studies using different incarnations of the bulb model, i.e., a stationary boundary with static solutions evolving as a function of radius. The later extension to time-dependent boundary conditions is conceptually similar in that a real frequency $\omega\not=0$ is assumed and one looks for real or complex $\bk$, i.e., spatial instabilities, not temporal ones. Which of these pictures, if any, better captures reality is a questions that has not been investigated yet. The answer may be provided by distinguishing more carefully between the character of the instabilities as absolute (local) or convective, where a growing perturbation moves away from the region where it was born. This question certainly depends on the underlying neutrino angular distribution that in a SN, depending on location, could be nearly isotropic or strongly beamed. We plan to address this second question of whether the instabilities will relax within the region where they were born in a follow-up work, based on the systematic understanding we develop here.

To develop a systematic understanding of the slow dispersion relation in an infinite homogeneous medium, we begin in Sec.~\ref{sec:EOMs} with a recap of the two-flavor equations of motion and derive the dispersion relation for a homogeneous but nonisotropic system that is monochromatic for neutrinos and antineutrinos. In Sec.~\ref{sec:homogeneous-isotropic} we review the homogeneous and isotropic case with homogeneous solutions only. Next we turn in Sec.~\ref{sec:non-resonant} to inhomogeneous modes, in a neutrino density range, where the instabilities are not resonant and thus somewhat resemble the homogeneous solutions. A different class of modes is studied in Sec.~\ref{sec:resonant}, the high-density regime, where the unstable modes derive from resonant interaction with individual neutrino modes. We explore the different regimes in a numerical example in Sec.~\ref{sec:numerical} with surprisingly complicated dispersion relations even for a benign neutrino and antineutrino angle distribution. We summarize our conclusions in Sec.~\ref{sec:conclusions}.

\clearpage

\section{Equations of motion}
\label{sec:EOMs}

In this section, we summarize our chosen setup and the equation of motion (EOM) for this case, forming the basis for our subsequent analysis of slow instabilities.

\subsection{Axially symmetric system}

We assume axial symmetry, reducing phase space to the variables
time $t$, spatial coordinate $r$, and neutrino velocity along that direction $v=\cos\theta$, where this latter choice of notation is taken from a related early paper \cite{Raffelt:2007yz}. We use the letter $r$ for the spatial coordinate to save $z$ for the weak-interaction direction in flavor space. However, $r$ should not necessarily suggest the radial direction in a SN, it is a general symmetry direction. The energy spectrum is taken to be monochromatic with energy $E$. The quantum kinetic equation \eqref{eq:QKE} implies the spatially one dimensional EOM
\begin{equation}
i(\partial_t+v\partial_r)\varrho(v,r,t)
=\bigl[\sH(v,r,t),\varrho(v,r,t)\bigr]
\end{equation}
and analogous for antineutrinos. We now neglect collisional interactions among neutrinos or with external matter, which might lead to novel branches of collisional instabilities~\cite{Johns:2021qby,Xiong:2022zqz, Liu:2023pjw, Lin:2022dek, Johns:2022yqy, Padilla-Gay:2022wck, Fiorillo:2023ajs}. The dimensionally reduced Hamiltonian matrices \eqref{eq:Hamiltonian} driving the evolution become
\begin{equation}
    \sH(v,r,t)=\pm\frac{\sM^2}{2E}
    +\sqrt{2}\GF\sN+\mu \int_{-1}^{+1}dv'\,\bigl[\varrho(v',r,t)-\bar\varrho(v',r,t)\bigr](1-v'v).
\end{equation}
The neutrino density matrices $\varrho$ are now taken to be integrated over most of phase space and normalized to ${\rm Tr}\int_{-1}^{+1}dv\,\varrho(v,r,t)=n_\nu/(n_\nu+n_{\overline{\nu}})$, the relative local number density of neutrinos of all flavors. This normalization makes the density matrices dimensionless, and the effective neutrino-neutrino interaction strength is \hbox{$\mu=\sqrt{2}\GF (n_\nu+n_{\overline{\nu}})$}, a parameter which in principle depends on space. However, we will ignore spatial gradients of overall physical parameters, assuming that the characteristic scales of collective instabilities are much smaller than overall geometric scales.

\subsection{Two-flavor case and mass ordering}

We restrict our discussion to two flavors, where it is convenient to express any Hermitian $2{\times}2$ matrix $\sA$ in terms of polarization vectors $(A_0,\vec{A})$ by virtue of $\sA=\frac{1}{2}(A_0\sigma_0+\vec{A}\cdot\vec{\sigma})$. Here, $\vec{\sigma}$ is a vector of Pauli matrices, $\sigma_0$ is the $2{\times}2$ unit matrix, $A_0={\rm Tr}\,\sA$, and $A_i={\rm Tr}(\sA\sigma_i)$ with $i=1,2,3$. In the flavor basis, the matter term is written as $\frac{1}{2}\lambda\sigma_3$ with $\lambda=\sqrt{2}\GF(n_{e^-}-n_{e^+})$ if only charged leptons of the electron flavor are present. Moreover, for the vacuum oscillation piece we write in the usual convention and notation
\begin{equation}
    \frac{\sM^2}{2E}=\frac{m_2^2+m_1^2}{2E}\frac{\sigma_0}{2}
    +\frac{m_2^2-m_1^2}{2E}\,\frac{\vec{B}\cdot\vec{\sigma}}{2},
\end{equation}
with $m_1<m_2$ being the masses of the two neutrino mass eigenstates. Here the ``magnetic field'' is a unit vector in flavor space which in the flavor basis has the components
\begin{equation}
    \vec{B}=(\sin2\theta_{\rm V},0,-\cos2\theta_{\rm V}),
\end{equation}
where $\theta_{\rm V}$ is the vacuum mixing angle. The vacuum oscillation frequency is denoted by
\begin{equation}
    \omegaE=\left|\frac{m_1^2-m_2^2}{2E}\right|
\end{equation}
and defined to be positive. Thus, the Hamiltonian engendering vacuum flavor evolution is
\begin{equation}
    {\sf H}_{\rm V}=\frac{\omegaE}{2}
    \begin{pmatrix}-\cos2\theta_{\rm V}& \sin2\theta_{\rm V}\\
    \sin2\theta_{\rm V}&\cos2\theta_{\rm V}\end{pmatrix}.
\end{equation}
Since $\omegaE$ is positive, $\cos2\theta_{\rm V}>0$ implies normal mass ordering, while $\cos2\theta_{\rm V}<0$ implies inverted ordering.

\subsection{EOMs in precession form}

We assume a system that is initially homogeneous, implying that neutrino densities are conserved, i.e., only the trace-free part of the density matrices evolves nontrivially under neutrino-neutrino and matter refraction. In this context one often uses the traditional polarization vectors $\vec{P}={\rm Tr}(\varrho\vec{\sigma})$ to express the density matrices in the form $\varrho-\frac{1}{2}{\rm Tr}(\varrho)=\frac{1}{2}\vec{P}\cdot\vec{\sigma}$. In particular, for the $z$-component this means $P_z=\varrho_{ee}-\varrho_{xx}$, where the second flavor is called~$x$. The usual precession form of the EOMs is
\begin{subequations}
    \begin{eqnarray}
       \kern-2em
        (\partial_t+v\partial_r)\vec{P}(v)&=&\left\{+\omegaE\vec{B}+\lambda\vec{L}
        +\mu\int_{-1}^{+1}dv'\left[\vec{P}(v')-\vec{\bar{P}}(v')\right](1-v'v)\right\}\times\vec{P}(v),
        \\
        \kern-2em
        (\partial_t+v\partial_r)\vec{\bar{P}}(v)&=&\left\{-\omegaE\vec{B}+\lambda\vec{L}
        +\mu\int_{-1}^{+1}dv'\left[\vec{P}(v')-\vec{\bar{P}}(v')\right](1-v'v)\right\}\times\vec{\bar{P}}(v),
    \end{eqnarray}
\end{subequations}
where the space-time dependence is no longer shown explicitly, but always assumed. Following previous notation, $\vec{L}$ is a unit vector in the flavor direction, identical with $\hat{\bf z}$, and $\lambda$ the precession caused by a homogeneous matter background.

Notice that we do {\em not} use the flavor isospin convention, i.e., if both neutrinos and antineutrinos are initially in the electron flavor, both $\vec{P}$ and $\vec{\bar P}$ initially point ``up'' in the positive $z$ direction in flavor space. Therefore, the polarization vector for flavor lepton number will be $\vec{D}=\vec{P}-\vec{\bar P}$.

These equations can be written more compactly if we introduce angular moments of the type 
\begin{equation}\label{eq:angular-moments}
\vec{P}_n=\int_{-1}^{+1}dv\,v^n\vec{P}(v)
\end{equation}
with $\vec{P}_0$ being the density and $\vec{P}_1$ the flux along the $r$ direction. The velocity $v'$ now disappears from the equations, being already integrated over. Therefore, for notational convenience, we no longer need to show the dependence of $\vec{P}$ on $v$ explicitly. With these simplifications one finds
\begin{subequations}\label{eq:EOM}
    \begin{eqnarray}
        (\partial_t+v\partial_r)\vec{P}&=&
        \left\{+\omegaE\vec{B}+\lambda\vec{L}
        +\mu\left[\bigl(\vec{P}_0-\vec{\bar{P}}_0\bigr)-v\bigl(\vec{P}_1-\vec{\bar{P}}_1\bigr)\right]\right\}\times\vec{P},
        \\
        (\partial_t+v\partial_r)\vec{\bar{P}}&=&
        \left\{-\omegaE\vec{B}+\lambda\vec{L}
        +\mu\left[\bigl(\vec{P}_0-\vec{\bar{P}}_0\bigr)-v\bigl(\vec{P}_1-\vec{\bar{P}}_1\bigr)\right]\right\}\times\vec{\bar{P}}.
    \end{eqnarray}
\end{subequations}
These equations become even more transparent if we introduce the sum and difference vectors, denoting particle number and lepton number, respectively, by $\vec{S}(v)=\vec{P}(v)+\vec{\bar{P}}(v)$ and $\vec{D}(v)=\vec{P}(v)-\vec{\bar{P}}(v)$, and analogous for the angular moments, leading to
\begin{subequations}
    \begin{eqnarray}
        (\partial_t+v\partial_r)\,\vec{S}&=&\omegaE\vec{B}\times\vec{D}
        +\bigl[\lambda\vec{L}+\mu(\vec{D}_0-v\vec{D}_1)\bigr]\times\vec{S},
        \\ \label{eq:D-EOM}
        (\partial_t+v\partial_r)\vec{D}&=&\omegaE\vec{B}\times\vec{S}\,
        +\bigl[\lambda\vec{L}+\mu(\vec{D}_0-v\vec{D}_1)\bigr]\times\vec{D}.
    \end{eqnarray}
\end{subequations}
The pure fast flavor case is defined by $\omegaE=0$, where the second equation becomes self-contained. The first equation is linear in the $\vec{S}$ variables and can be integrated once the equation for the $\vec{D}$ variables has been solved.

In the fast flavor case, an instability with wave vector along the axis of symmetry is certain to appear when the angular lepton-number spectrum $D^z(v)|_{t=0}$ has a single crossing~\cite{Capozzi:2019lso, Fiorillo:2024bzm, Fiorillo:2024uki}, i.e., it changes sign once at some value of $v$. Notice that this statement is different from, and generally not implied by, Morinaga's theorem~\cite{Morinaga:2021vmc, Fiorillo:2024bzm}, which states that if the distribution has any angular crossing (even more than one) there will be some unstable modes. However, they need not be directed along the symmetry axis, and therefore need not appear in a one-dimensional formulation. 

If there is no angular crossing, as we assume here, there are no unstable modes in the limit $\omegaE\to 0$. Therefore, any instability for $\omegaE\neq 0$ is slow according to the conventional definition, i.e., must vanish as $\omegaE\to 0$. We will show here that the properties of such instabilities are however much less universal than often assumed; the precise way the growth rate vanishes, the length scales of the unstable modes, and the impact these modes have on the angular distribution, can have markedly different characteristics from what is sometimes implied in the literature.

\subsection{Linearization}

As usual in this context, we assume that the neutrinos are initially in flavor eigenstates so that $\vec{P}(v)$ and $\vec{\bar P}(v)$ are nearly aligned with the $z$ axis, the flavor direction. Therefore, we will treat the transverse components of the polarization vectors $P^x$ and $P^y$ as small perturbations, whereas $P^z$ remains fixed at its initial value that we call the spectrum. More specifically, hopefully without causing notational confusion, we will use
\begin{subequations}
\begin{eqnarray}
    P(v)&=&P^z(v)|_{t=0}\kern2.25em\hbox{Angular spectrum for neutrinos,}\\
    \oP(v)&=&\oP^z(v)|_{t=0}\kern2.25em\hbox{Angular spectrum for antineutrinos,}\\
    D(v)&=&P(v)-\oP(v)\quad\hbox{Angular spectrum for lepton number,}\\
    S(v)&=&P(v)+\oP(v)\quad\hbox{Angular spectrum for particle number,}
\end{eqnarray}    
\end{subequations}
where $P(v)>0$ and $\oP(v)>0$ if both neutrinos and antineutrinos begin in the electron flavor.
Since the perturbed motion is assumed to be completely transverse to the flavor axis, it is fully described by a single complex variable $\psi(v)=P^x(v)+iP^y(v)$. Linearizing the EOM in $\psi$, we find
\begin{subequations}\label{eq:eom_linearized_inhomogeneous}
    \begin{eqnarray}
        \kern-2.5em
        (\partial_t+v\partial_r)\psi&=&
        -i(\omega_E \cV-\lambda)\psi+i\mu\bigl[\,\psi (D_0-v D_1)-P (\Psi_0-v \Psi_1)\bigr] -i\omega_E \sV P,
        \\
        \kern-2.5em
        (\partial_t+v\partial_r)\opsi&=&
        +i(\omega_E\cV+\lambda)\opsi +i\mu\bigl[\,\opsi (D_0-v D_1)-\oP (\Psi_0-v \Psi_1)\bigr] +i\omega_E \sV\oP,
    \end{eqnarray}
\end{subequations}
where we have introduced the notation $\Psi(v)=\psi(v)-\opsi(v)$ for the lepton-number field of flavor coherence.  Moreover, we use angular moments of the type Eq.~\eqref{eq:angular-moments}. The vacuum mixing angle enters through $\cV=\cos 2\theta_{\rm V}$ and $\sV=\sin 2\theta_{\rm V}$ and we take the mixing angle to lie in the interval $0\leq \theta_{\rm V}\leq\pi/2$, meaning that the octant $\pi/4< \theta_{\rm V}\leq\pi/2$, where $\cV<0$, corresponds to inverted mass ordering.

These equations can be grouped into a homogeneous system, with a source term on the right-hand side proportional to $\omega_E \sV$, which therefore acts as the primary perturbation triggering the motion~\cite{Airen:2018nvp}. We can also regard this term as an external field acting on the system and potentially triggering its instabilities~\cite{Fiorillo:2024bzm}. Linear stability analysis corresponds to asking the question: does the homogeneous part of this system admit exponentially growing solutions? If the answer is yes, then one can decompose the solution of the full inhomogeneous system of equations into normal modes of the homogeneous system, plus a particular solution of the inhomogeneous system, to obtain the solution to the initial-value problem, where the initial value corresponds to $\psi(t=0)=\overline{\psi}(t=0)=0$. This strategy is followed for a discrete set of neutrino beams in Ref.~\cite{Airen:2018nvp}. A direct solution can also be found by applying a Laplace transform to the equations, as we did in Ref.~\cite{Fiorillo:2024bzm}. The two approaches lead to a similar conclusion, namely that after an initial transient phase the transverse components will asymptotically grow if the homogeneous system possesses unstable eigenmodes, while they will remain small if no such eigenmode exists. We are thus motivated to continue our linear stability analysis by looking for the normal modes of the homogeneous system. 

To determine these normal modes, we seek a solution $\psi\to \psi e^{-i\Omega t+i K r}$, where we are only considering solutions with the wave vector along the symmetry axis, where 
$K$ is the corresponding wave number. Moreover, we introduce the shifted variables $\omega=\Omega+ \mu D_0+\lambda$ and $k=K+ \mu D_1$ so that the solution is of the form
\begin{subequations}\label{eq:longitudinal_eigenmode}
    \begin{eqnarray}
        \psi&=&\frac{P}{\omega-k v -\tomegaE}(\Psi_0-v \Psi_1),
        \\
        \opsi&=&\frac{\oP}{\omega-k v +\tomegaE}(\Psi_0-v \Psi_1).
    \end{eqnarray}
\end{subequations}
Here we have set $\mu=1$ by a redefinition of the scales of space and time and we have introduced the more compact notation $\tomegaE=\omegaE \cos 2\theta_{\rm V}$, a quantity that is positive for normal mass ordering and negative for inverted ordering. The matter term $\lambda$ has disappeared from the equations, amounting merely to a shift in the real part of the eigenfrequency $\omega$. Yet this shift has direct physical consequences, as we discuss later in Sec.~\ref{sec:large_mixing_angles}.

Inserting these forms of the solution into the homogeneous form of the EOM~\eqref{eq:eom_linearized_inhomogeneous} provides the self-consistency condition, or dispersion relation, 
\begin{equation}\label{eq:dispersion}
    (\tilde I_0-1)(\tilde I_2+1)-\tilde I_1^2=0,
\end{equation}
where we use
\begin{equation}\label{eq:def_In}
    \tilde I_n=\int_{-1}^{+1}dv \frac{P(v) v^n}{\omega-k v-\tomegaE}-\int_{-1}^{+1}dv \frac{ \oP(v) v^n}{\omega-k v+\tomegaE}.
\end{equation}
We use the notation $\tilde I_n$ for the dispersion relation based on normal modes, whereas we will use $I_n$ for the physical modes that require Landau's $i\epsilon$ prescription as discussed later.

So far we have assumed that $\psi$ depends only on $v$, and not on the azimuth angle $\phi$ around the symmetry axis. Such axial-symmetry-breaking modes exist even if the unperturbed distributions $P(v)$ and $\oP(v)$ are azimuthally symmetric. They can be found by noting that for them the quantity $\psi_1\to \bpsi_1=\int_{-1}^{+1} dv \int_0^{2\pi}\frac{d\phi}{2\pi} \sum_{\tomegaE} \psi(v,\phi) \bv$, where $\bv=(\sqrt{1-v^2}\cos\phi,\sqrt{1-v^2}\sin\phi,v)$ is the velocity vector. Therefore, the dispersion relation is generalized to 
\begin{equation}\label{eq:general_eigenmode}
    \psi(v,\phi)=\frac{P(v)}{\omega-kv-\tomegaE}(\psi_0-\bv\cdot\bpsi_1)
\end{equation}
and analogous for antineutrinos with $\oP(v)$ and $-\tomegaE$. The axial symmetry breaking solutions are found by assuming $\psi(v,\phi)\propto \cos\phi$ or $\sin\phi$, which are degenerate, so that $\psi_0=0$ and the consistency condition after multiplying by $\cos\phi$ or $\sin\phi$ respectively becomes
\begin{equation}\label{eq:dispersion_transverse}
    \tilde I_0-\tilde I_2+2=0.
\end{equation}
Relations~\eqref{eq:dispersion} and~\eqref{eq:dispersion_transverse} form the basis for our analysis of slow unstable modes. 

So far, we have discussed the normal modes of the homogeneous system with the goal of finding the unstable (exponentially growing) ones. On the other hand, in Ref.~\cite{Fiorillo:2024bzm} we showed that, if one solves the full initial-value problem including the inhomogeneous source term, the behavior of the system at late times is given by effective modes with eigenfrequencies coming from a modified dispersion relation. The reason is that, while unstable eigenmodes are correct normal modes of the system, when we have a continuum of velocities for the neutrinos we can also have an asymptotic behavior where the perturbation is damped in time because of decoherence among different velocity modes. This form of damping, known as Landau damping, is reversible, since there is no scattering process involved in the equations, and does not correspond to a normal mode of the system.

The modified dispersion relation is formally identical to the one derived above, but the integrals $\tilde I_n$ must be modified to
\begin{equation}\label{eq:def_In-Landau}
    {I}_n=\int_{-1}^{+1}dv \frac{ P(v) v^n}{\omega-k v-\tomegaE+i\epsilon}-\int_{-1}^{+1}dv \frac{\oP(v) v^n}{\omega-k v+\tomegaE+i\epsilon}.
\end{equation}
Landau's $i \epsilon$ prescription introduced here ensures causality, i.e., the eigenfrequencies from the dispersion relation are the ones actually describing the evolution into the far future of the system. From the practical point of view, the introduction of this term means that the integration contour along the $v$ variable from $-1$ to $+1$ should also be deformed so as to pass below the singular point of the denominator $\overline{v}=(\omega\pm \tomegaE)/k$. Neutrino modes matching this Cherenkov condition can resonantly extract or deposit energy, leading to Landau damping or instability. We will later use this modified form of the dispersion relation to obtain the collective solutions which include the Landau-damped branches, since only in this way the analytical properties of the system are easily tractable.

\subsection{Large mixing angles and matter effect}\label{sec:large_mixing_angles}

So far, we have treated the term proportional to $\sin2\theta_{\rm V}$ in Eq.~\eqref{eq:eom_linearized_inhomogeneous} as a source term, and we have explicitly stated that the asymptotic behavior at late times is entirely independent of this term, which acts only as a normalization for the induced perturbation; the question of whether such perturbation grows or not has been completely linked with the normal modes of the homogeneous system. Is this always the case? The answer to this question cannot be yes, since we are in fact well aware of cases where this is not true. The simplest example consists of vacuum oscillations with maximal mixing ($\theta_{\rm V}=\pi/4$). Neutrinos beginning in the $\nu_e$ state completely convert to the other flavor and back and therefore large flavor coherence develops without any collective instability. On the linear level, the EOM in this case ($\mu=\lambda=k=0$) reduces to
\begin{equation}\label{eq:first_example}
    \partial_t \psi=-i\omega_E \sin 2\theta_{\rm V} P,
\end{equation}
showing that $\psi$ grows linearly with time even though the homogeneous system is obviously stable, and similar for $\overline\psi$.

As a slightly less trivial example, we now include neutrino-neutrino refraction $\mu$ (which we temporarily restore explicitly in the equations), but still ignore the neutrino-matter term $\lambda$, i.e., we consider the well-known flavor pendulum \cite{Hannestad:2006nj}, but with the unusual choice of maximal mixing. We assume a single velocity mode $v$, so that $\Psi_1=v\Psi_0=v(\psi-\opsi)$, and similarly for $P$ and $\oP$, so that the linear EOMs become
\begin{subequations}\label{eq:second_example}
\begin{eqnarray}
    \partial_t\psi&=&i \mu(1-v^2) \psi D-i\mu (1-v^2) P(\psi-\opsi)-i\omega_E\sin2\tV P, 
    \\
    \partial_t \opsi&=&i \mu (1-v^2) \overline{\psi} D-i\mu (1-v^2) \oP(\psi-\opsi)+i\omega_E\sin2\tV \oP.
\end{eqnarray}    
\end{subequations}
If we take the difference of these equations, the terms proportional to $\mu$ drop out and what remains is $\partial_t(\psi-\opsi)=-i\omega_E \sin2\tV (P-\oP)$ so that $\psi-\opsi$ grows linearly in time.

These simple examples show that there can be power-law growth induced by the term proportional to $\sin2\tV$ even when linear stability analysis does not predict unstable eigenmodes. However, these examples also reveal the conditions for this to happen, namely that the homogeneous system of equations should have a mode with zero eigenfrequency. Indeed, the homogeneous form of Eq.~\eqref{eq:first_example} is $\partial_t \psi=0$ and therefore has zero eigenfrequency. Likewise, the homogeneous form of the difference of Eqs.~\eqref{eq:second_example} is $\partial_t(\psi-\overline{\psi})=0$, which again corresponds to a zero eigenfrequency. If the system does not possess normal modes with vanishing eigenfrequencies (or more exactly with eigenfrequencies of order $\omega_E$), this secular growth of perturbations is impossible. 

Actually, this fact is well known, although here somewhat masked in the formalism of linear stability analysis. To see why this is the case, let us consider again the example of a single, non self-interacting neutrino beam from Eq.~\eqref{eq:first_example}, but now introduce a matter effect, so that
\begin{equation}
    \partial_t \psi=i\lambda \psi-i\omega_E \sin2\tV P
\end{equation}
with the explicit solution 
\begin{equation}
    \psi=\frac{\omega_E \sin 2\tV P}{\lambda}\bigl(1-e^{i\lambda t}\bigr).
\end{equation}
We now see that if $\lambda\gg \omega_E$, the perturbation always remains small; large matter refraction precludes non-collective forms of flavor conversion, as well known. On the other hand, if $\lambda\lesssim \omega_E$, for times of order $t\sim \omega_E^{-1}$, we may expand the exponential in parenthesis and recover the linear growth, so the perturbation grows to become of order $\psi\sim\sin2\tV P$. 

Therefore, our main takeaway is that, if the frequencies of the homogeneous system satisfy the condition $\omega \gg \omega_E$, then the only possible growth of the perturbation comes from unstable normal modes. This insight justifies our use of linear stability analysis throughout the text. The possibility of having normal modes with frequency $\omega\lesssim \omega_E$ in the full system is possible but fine tuned, as the examples above show. In the case of the flavor pendulum, the existence of a zero-frequency eigenmode is guaranteed by the law of conservation of lepton number; by integrating Eq.~\eqref{eq:D-EOM} over $v$ we find
\begin{equation}
    \partial_t \vec{D}_0+\partial_r \vec{D}_1=\omega_E\vec{B}\times \vec{S}_0+\lambda \vec{L}\times \vec{D}_0.
\end{equation}
Thus, if $\lambda=0$ and if we focus only on homogeneous modes, we see that the vector $\vec{D}_0$ changes only over timescales of order $\omega_E^{-1}$, thus very slowly. However, once matter refraction is introduced, and once a degree of inhomogeneity $K\sim \mu$ is considered, this protected slow variation disappears and all the modes change over timescales much shorter than $\omega_E^{-1}$. In the presence of inhomogeneities, matter, and anisotropies, the possibility of having fine-tuned situations with vanishing eigenfrequencies essentially disappears. The flavor pendulum is a unique example, illustrating just how fragile this possibility is, since it requires perfect homogeneity and isotropy.

Another comment pertains to the argument used at times that matter refraction $\lambda$ can be removed in exchange for introducing an effectively small mixing angle. As we have discussed, this is generally not true. The unstable modes entering the dispersion relation are determined by $\tomegaE=\omega_E \cos2\tV$, the projected eigenfrequency, where $\tV$ is the \textit{vacuum} mixing angle. In the same way, the forcing term in Eq.~\eqref{eq:eom_linearized_inhomogeneous} is proportional to $\omega_E \sin2\tV$, again depending on the vacuum mixing angle. Matter will suppress the amplitude of the induced perturbation, but this must be done self-consistently by solving the inhomogeneous system, and not by mimicking its effect through a small effective mixing angle. First, this procedure is not predictive, since one cannot in general know in advance how much matter will suppress the amplitude of the conversions, which depends on the precise solution of Eq.~\eqref{eq:eom_linearized_inhomogeneous}.  Second, the unstable eigenmodes depend on $\tomegaE$, and using a suppressed mixing angle would therefore incorrectly reproduce the growth rate of these eigenmodes; since this growth rate appears in the exponent, this would end up dramatically changing the solution if $\tV$ is not already very small.

\section{Homogeneous and isotropic neutrino gas}
\label{sec:homogeneous-isotropic}

In this section, we will review the instability pattern of a homogeneous and isotropic neutrino gas, focusing on homogeneous instabilities ($K=0$). One of them, the slow flavor pendulum that occurs for inverted mass ordering, is well known in the literature \cite{Samuel:1993uw, Hannestad:2006nj, Duan:2007mv, Raffelt:2011yb}. Another branch that occurs for normal mass ordering was identified a long time ago \cite{Raffelt:2007yz}, but is much less familiar. In any case, a systematic discussion of these instabilities in modern language seems to be lacking, so we summarize here the main properties of what is the simplest neutrino system to exhibit an instability.

\subsection{Dispersion relation}

In the homogeneous and isotropic case we have $D_1=0$ and therefore $K=k=0$, implying that all the integrals $\tilde I_n$ become simple functions of the moments of the angular distributions
\begin{equation}
    \tilde I_n=\frac{P_n}{\omega-\tomegaE}-\frac{\oP_n}{\omega+\tomegaE}.
\end{equation}
Moreover, also for the separate $\nu$ and $\bar\nu$ distributions we have $P_1=\oP_1=0$, implying $\tilde I_1=0$. Therefore, the dispersion relation for longitudinal modes Eq.~\eqref{eq:dispersion} factorizes as $(\tilde I_0-1)(\tilde I_2+1)=0$ or explicitly
\begin{equation}\label{eq:factorized-dispersion-relation}
   \tilde I_0=+1
   \quad\hbox{or}\quad
   \tilde I_2=-1.
\end{equation}
Moreover, isotropy implies $P_2=P_0/3$ and $\oP_2=\oP_0/3$ and thus
$\tilde I_2=\tilde I_0/3$ so that Eq.~\eqref{eq:factorized-dispersion-relation} eventually falls into two families of solutions with
\begin{subequations}
    \begin{eqnarray}\label{eq:dispersion_monopole}
        \tilde I_0(\omega)&=&+1,
        \\ \label{eq:dispersion_dipole_z}
        \tilde I_0(\omega)&=&-3,
    \end{eqnarray}
\end{subequations}
which we call monopole and dipole, respectively. To understand the meaning of this terminology, we notice that the monopole mode is obtained assuming $\psi(v)$ independent of $v$, so that $\psi_1=0$ and Eq.~\eqref{eq:longitudinal_eigenmode} immediately leads to Eq.~\eqref{eq:dispersion_monopole}. If instead we assume $\psi(v) \propto v$, we have $\psi_0=0$; after multiplying Eq.~\eqref{eq:longitudinal_eigenmode} by $v$ and integrating, we find Eq.~\eqref{eq:dispersion_dipole_z}. 

In addition, for the transverse modes obeying Eq.~\eqref{eq:dispersion_transverse} we find $\tilde I_0=-3$, exactly as for the dipole mode. Indeed, these transverse modes are also of the dipole form; for them, we have $\psi(v,\phi)\propto \sqrt{1-v^2}\cos\phi$ or $\sqrt{1-v^2}\sin\phi$, so they are the same dipole mode oriented along different axes. It is of course obvious that if the background distribution is homogeneous and isotropic, its eigenmodes must be defined by spherical harmonics. So we conclude that this simple problem admits one monopole and three degenerate dipole eigenmodes.

\subsection{Monopole mode}

To determine whether one of these modes can become unstable, we proceed to find explicitly their eigenfrequencies. For the monopole mode, we find from Eq.~\ref{eq:dispersion_monopole}
\begin{equation}
    \omega=\frac{P_0-\oP_0}{2}\pm\frac{\sqrt{(P_0-\oP_0)^2+4 (P_0+\oP_0)\tomegaE+4\tomegaE^2}}{2}.
\end{equation}
In our normalization, where $\mu=\sqrt{2}\GF(n_\nu+n_{\bar\nu})$, we effectively have $P_0+\oP_0=1$ as well as $\epsilon=(P_0-\oP_0)/(P_0+\oP_0)=(P_0-\oP_0)$ for the neutrino-antineutrino asymmetry introduced earlier. After restoring $\mu$ explicitly, we thus find
\begin{equation}
    \omega=\frac{\epsilon\mu}{2}\pm\frac{\sqrt{(\epsilon\mu)^2+4\mu\tomegaE+4\tomegaE^2}}{2}.
\end{equation}
By our convention, $\tomegaE>0$ for normal mass ordering, in which case the discriminant under the square root is always positive and the two eigenfrequencies $\omega$ are real. Conversely, $\tomegaE<0$ for inverted mass ordering and the discriminant is negative for
\begin{equation}
    -\frac{\mu}{2}\left(1+\sqrt{1-\epsilon^2}\right)<\tomegaE< -\frac{\mu}{2}\left(1-\sqrt{1-\epsilon^2}\right),
\end{equation}
implying two complex conjugate eigenfrequencies $\omega$. In practice, we are interested in a dense neutrino gas with $\mu\gg|\tomegaE|$ so that only the second inequality is relevant which implies a maximal $\mu$ for fixed $\tomegaE<0$ to have an instability. If the asymmetry parameter $\epsilon$ is not too large, we may expand the square root and the condition for instability is $\frac{1}{4}\epsilon^2\mu<|\tomegaE|$, which for small $\epsilon$ is consistent with the requirement $|\tomegaE|\ll\mu$. In this case we may neglect $\tomegaE^2$ under the square root and the growth rate is
\begin{equation}
    {\rm Im}\,\omega=\sqrt{\mu\left(|\tomegaE|-\frac{\epsilon^2\mu}{4}\right)}.
\end{equation}
Therefore, the threshold value for $|\tomegaE|$ is $\epsilon^2\mu/4$ and for significantly larger values, the growth rate is $\mathrm{Im}(\omega)\simeq\sqrt{|\tomegaE| \mu}$, leading to the characteristic scaling sometimes identified as symptomatic of slow flavor conversions \cite{Tamborra:2020cul, Sen:2024fxa}. As we will see, however, this is not generally true for inhomogeneous modes, and even in the simplest case of this homogeneous and isotropic system, this scaling applies only for $|\tomegaE|\gg \epsilon^2\mu/4$.

To complete the discussion of the monopole instability, we mention that in the nonlinear regime it continues as a regular periodic solution with a dynamical behavior analogous to a spherical gyroscopic pendulum or spinning top \cite{Samuel:1993uw, Hannestad:2006nj, Duan:2007mv}. Underlying this remarkable behavior is that for a perfectly homogeneous and isotropic neutrino gas, and only for the monopole instability -- so the initial perturbation is isotropic -- the system shows an infinity of conservation laws, the Gaudin invariants, and thus is technically integrable \cite{Pehlivan:2011hp, Raffelt:2011yb}. Actually, this ``slow flavor system'' can be mapped on an equivalent fast flavor system that is homogeneous but anisotropic, the fast flavor pendulum \cite{Johns:2019izj, Padilla-Gay:2021haz}, having analogous Gaudin invariants~\cite{Fiorillo:2023hlk}. However, we stress that this regular dynamics, as well as the corresponding flavor soliton \cite{Fiorillo:2023mze}, is extremely fragile. Any deviation from perfect symmetry (slight inhomogeneities, inhomogeneous perturbations, slight anisotropy) will break the regularity of the nonlinear dynamics and ultimately lead to some form of decoherence and flavor turbulence~\cite{Raffelt:2007yz, Mangano:2014zda}. Moreover, collisions will damp the pendular motion \cite{Padilla-Gay:2022wck, Fiorillo:2023ajs}. Therefore, the regular pendulum behavior is never expected to arise in a realistic system.

\subsection{Dipole mode}

Let us now turn to the dipole mode that has the dispersion relation given by Eq.~\eqref{eq:dispersion_dipole_z}, providing the eigenfrequency
\begin{eqnarray}
    \omega&=&-\frac{P_0-\oP_0}{6}\pm\frac{\sqrt{(P_0-\oP_0)^2-12 (P_0+\oP_0)\tomegaE+36\tomegaE^2}}{6}
    \nonumber\\
    &=&-\frac{\epsilon\mu}{6}\pm\frac{\sqrt{(\epsilon\mu)^2-12\mu\tomegaE+36\tomegaE^2}}{6},
\end{eqnarray}
where in the second line we have once more restored $\mu$. For negative $\tomegaE$ (inverted mass ordering), the discriminant is positive, there are two real eigenfrequencies, and no instability. Conversely, for positive $\tomegaE$ (normal ordering), the discriminant is negative for
\begin{equation}
    \frac{\mu}{6}\left(1-\sqrt{1-\epsilon^2}\right)
    <\tomegaE<
    \frac{\mu}{6}\left(1+\sqrt{1-\epsilon^2}\right).
\end{equation}
Expanding once more in powers of small $\epsilon$ implies $\frac{1}{12}\epsilon^2\mu<\tomegaE$ as a requirement for instability and in the limit $\tomegaE\ll\mu$ the growth rate is
\begin{equation}
    {\rm Im}\,\omega=\sqrt{\frac{\mu}{3}\left(\tomegaE-\frac{\epsilon^2\mu}{12}\right)}.
\end{equation}
Being intrinsically anisotropic, the dipole instability never leads to a regular or periodic behavior~\cite{Raffelt:2007yz}.

\subsection{Final remarks}

One further feature of these instabilities relevant for their physical interpretation is that under the integral the denominator $\omega-k v\pm\tomegaE$  does not vanish for any value of $v$ because $k=0$. In terms of the plasma-physics analogy that we have introduced earlier \cite{Fiorillo:2024qbl, Fiorillo:2024uki}, we say these instabilities are non-resonant. In contrast, resonant instabilities grow out of the interaction with specific neutrino modes for which the denominator vanishes, but for $k\to 0$, this would require neutrino modes with infinite velocity that do not exist. As we will see, the non-resonant nature of the instabilities will allow us to generalize their main features to generally anisotropic neutrino systems.

The main message of this simple homogeneous and isotropic example is that instabilities with large length scales, in the limit studied here homogeneous, appear in normal mass ordering, provided that $\frac{1}{12}\epsilon^2\mu<\tomegaE$, and in inverted ordering for $\tomegaE<-\frac{1}{4}\epsilon^2\mu$, always assuming $|\tomegaE|\ll\mu$. In normal ordering (positive $\tomegaE$), the unstable modes are three degenerate dipole modes, whereas the monopole mode is stable. For inverted ordering (negative $\tomegaE$), it is the other way around and the monopole mode is unstable, the usual slow flavor pendulum, whereas the three degenerate dipole modes are stable. The typical growth rate is of the order of $\sqrt{|\tomegaE|\mu}$ with different coefficients in both cases. We notice, though, that the real part of the eigenfrequency is actually of the order of $\epsilon\mu \ll \sqrt{|\tomegaE| \mu}$, i.e., of the same order as the typical frequencies of fast modes. In the next section, we will explore this regime of vacuum frequencies in more general terms, without restricting to the background distribution being necessarily isotropic.

\section{Non-resonant slow instabilities}\label{sec:non-resonant}

As we have discovered in the previous section, the well-known slow pendulum instability is a special example of a superluminal, non-resonant instability, for the case of a homogeneous and isotropic neutrino gas. Let us now consider generic angular distributions $P(v)$ and $\oP(v)$ that we still take to be axially symmetric and restrict ourselves to the regime \hbox{$\mu\epsilon^2\ll |\tomegaE|\ll \mu$}. Moreover, we now consider generically inhomogeneous modes with wavenumber $k=K+D_1$. Can we generalize some of the features of the homogeneous and isotropic example studied earlier? We will see that indeed many features of the instability stay unchanged in this regime.

\subsection{Small-flux expansion}

Motivated by the homogeneous case, we first consider wavenumbers of the order of $k\simeq \mu\epsilon$ and assume that $|\omega|\gg k$, which was certainly the case in the previous example since $\mathrm{Im}(\omega)\simeq \sqrt{|\tomegaE|\mu} \gg \mu \epsilon$. If $|\omega|\gg k$, we are assuming the instability to be strongly non-resonant. We may therefore safely expand the denominators
\begin{equation}
    \frac{1}{\omega-k v\pm \tomegaE}=\frac{1}{\omega\pm\omegaE}+\frac{kv}{(\omega\pm\omegaE)^2}+\frac{k^2 v^2}{(\omega\pm\omegaE)^3}+...
\end{equation}
Thus, each of the integrals $\tilde I_n$ can be expanded as
\begin{equation}
    \tilde I_n=\sum_{m=0}^{\infty} k^m \left[\frac{P_{n+m}}{(\omega-\tomegaE)^{1+m}}-\frac{\oP_{n+m}}{(\omega+\tomegaE)^{1+m}}\right].
\end{equation}
Close to $k=0$, we can keep only the lowest orders of the expansion, so that the dispersion relation becomes an algebraic one which can be solved explicitly in terms of the moments of the angular distributions. 

This property is analogous to what we found earlier for fast instabilities~\cite{Fiorillo:2024uki}, and generally descends from non-resonant instabilities interacting simultaneously with the entire angular distribution, so its global properties measured by the moments are sufficient to obtain the instability properties. The algebraic equation that results even for $k=0$ is quartic in $\omega$, and therefore its explicit solution is usually too complicated to be particularly illuminating. For distributions that are not too anisotropic, for which $P_1\ll P_0, P_2$, and similarly for antineutrinos, one may initially neglect $P_1$ and $\oP_1$ so that the quartic equation decouples again into the two equations $\tilde I_0=1$ and $\tilde I_2=-1$, as for the isotropic case shown in Eq.~\eqref{eq:factorized-dispersion-relation}.

The effect of the small flux encoded in $P_1$ and $\oP_1$ can then be incorporated perturbatively. We do not perform this calculation explicitly here, since it is not particularly instructive; we only note that the unperturbed solutions for $P_1\to 0$ and $\oP_1\to 0$ are guaranteed to exhibit an instability, either for the near-monopole mode with $\tilde I_0=1$ in inverted ordering, or for the near-dipole mode with $\tilde I_2=0$ in normal ordering. Of course, as soon as there is a small flux, the flavor pendulum will quickly decohere \cite{Raffelt:2007yz}. Still, on the linear level, the majority of the properties of the non-resonant instabilities are inherited from the homogeneous and isotropic case, which acts as some sort of prototype. 

\subsection{Large-scale slow modes are fast}

As in the homogeneous and isotropic case, $\mathrm{Re}(\omega)$ is generally expected to be of the order of $\mu \epsilon \ll \mathrm{Im}(\omega)$. We have not excluded the possibility that our anisotropic distributions could produce a crossing of $D(v)$ and thus imply the emergence of fast-unstable modes. While we do not consider explicitly this case in detail here, the growth rate for these modes would be of the order of $\mu \epsilon\ll \sqrt{|\tomegaE|\mu}$. Therefore, in the regime in which large-scale modes are slow unstable, requiring $|\tomegaE|\gg \mu \epsilon^2$, the slow modes are actually \textit{faster} than the fast modes. 

The often-made argument that slow modes are slower because $\sqrt{|\tomegaE| \mu}\ll \mu$ misses the point that fast modes have a growth rate of the order of $\mu \epsilon$ and that the scaling for slow modes $\sqrt{|\tomegaE| \mu}$ is \textit{only} valid for $|\tomegaE|\gg \mu \epsilon^2$. The corrected version of this argument shows that it is slow modes that are the fastest-growing ones in this regime. On the other hand, whether this regime has any relevance for realistic environments in SN cores is a different question that we do not tackle here. We only mention here that in the neutrino decoupling region of a SN core, $\mu$ is on the order of $10^5$~km$^{-1}$, whereas $\tomegaE\simeq 0.4$~km$^{-1}$. Therefore, this regime would require a fractional difference between the $\nu_e$ and $\overline{\nu}_e$ angular distributions of the order of $\epsilon\sim \sqrt{\tomegaE/\mu}\sim 0.1\%$ and thus unnaturally small.

Finally, we should stress the regime of applicability of our conclusions. We have so far assumed small deviations from isotropy, i.e., $P_1\lesssim P_0$ and $P_2$ and similarly for antineutrinos. In this case, a monopole and dipole mode can still be defined in an approximate way, and they maintain similar conditions of stability as the isotropic case. Thus, for this case, $\epsilon$ can still be used approximately as the asymmetry in the zeroth moment, and for $\tomegaE\gg \mu \epsilon^2$ a slow instability appears. This conclusion is true even for a distribution with an angular crossing, provided that the latter does not cause too large an anisotropy. On the other hand, if the angular distribution develops a large anisotropy, with $P_1\gtrsim P_0$ and $P_2$ and similarly for antineutrinos, the concept of monopole and dipole mode loses meaning, and one should return to an explicit solution of Eq.~\eqref{eq:dispersion} for $k=0$. In this case, the definition of $\epsilon$ as the asymmetry in the zeroth moments of the angular distribution is not helpful, since the other moments are involved as well, and $\tomegaE\gg \mu \epsilon^2$ does not generally guarantee non-resonant instabilities. We do not consider this case further here.

\section{Resonant slow instabilities}\label{sec:resonant}

In this section, we consider the regime opposite to the previous section, namely when $|\tomegaE|\ll \mu\epsilon^2$. This is the regime of large neutrino density, where the slow flavor pendulum would be stable even in the inverted position (``sleeping top'' regime), whereas inhomogeneities can introduce unstable modes. Our approach will be completely analytical, and we will validate our conclusions by a numerical analysis of a specific example in Sec.~\ref{sec:numerical}. As before, we assume that both the neutrino and antineutrino angle distributions $P(v)$ and $\oP(v)$ are positive everywhere and also assume that there is no angular crossing of $D(v)=P(v)-\oP(v)$, which is also taken to be positive. Therefore, in the limit $\tomegaE\to 0$ there is no fast instability. We stress that, since these instabilities will turn out to be resonant, depending on the angular distribution evaluated at specific directions, the definition of $\epsilon$ as the asymmetry in the zeroth moments of the angular distribution is not helpful. Rather, with $\epsilon$ we will mean the order of magnitude of the asymmetry between $P_v$ and $\oP_v$ across the entire angular range, assuming that the distribution is reasonably regular.

Since $\tomegaE$ is smaller than any other scale in the problem, one might expect its impact to be perturbative, in the sense of weakly renormalizing the properties of the modes for $\tomegaE=0$. This conclusion is, however, too naive and ultimately wrong. It turns out that an infinitesimally small $\tomegaE$ can by itself alter dramatically the properties of some of the modes. The catch to the argument of perturbativity is that, in the limit $\tomegaE=0$, there are some modes with a phase velocity arbitrarily close to the speed of light (near luminal). For these modes, an infinitesimally small perturbation is enough to change completely their properties, and even turn them from stable to unstable.

\subsection{Resonance cones}
\label{sec:resonancecones}

Before proceeding with our mathematical treatment, let us notice a fundamental physical difference to the case of fast instabilities, where neutrinos can resonate with flavor waves only when $\omega=k v$, where $-1\leq v\leq+1$ is the neutrino velocity along the symmetry axis. Hence there is a resonance cone, delimited by $\omega=\pm k$, in which flavor waves can resonate with neutrinos, namely subluminal waves. Instead, in the slow case, a wave can resonate with neutrinos if $\omega=kv+\tomegaE$, while it can resonate with antineutrinos if $\omega=kv-\tomegaE$. These conditions differ, opening the possibility of an instability even for very small~$\tomegaE$. 

We thus need to introduce two resonance cones, delimited by $\omega=\pm k+\tomegaE$ and $\omega=\pm k-\tomegaE$, for neutrinos and antineutrinos, respectively. These two cones no longer coincide with the light cone, but are slightly shifted up or down. This structure introduces a much richer phenomenology than for fast flavor waves, for which a wave could either resonate with neutrino modes, if it is subluminal, or not resonate, if it is superluminal. In the slow case, new possibilities open because a mode may resonate with only neutrinos or only antineutrinos. We will follow this trail of physical ideas to prove mathematically the existence of an instability in the limit of small $\tomegaE$.

\subsection{Near-luminal modes}

Motivated by these arguments, we will consider the impact of $|\tomegaE|\ll \mu \epsilon^2$ only on near-luminal modes, for which $\omega\simeq \pm k+\chi$, where $\chi$ is a small number. We first consider the case with $+k$. The integrals $I_n$ defined in Eq.~\eqref{eq:def_In} can then be expanded close to the luminal sphere by the procedure introduced in Ref.~\cite{Fiorillo:2024bzm}
\begin{equation}
    I_n=\frac{P(1)\log\left(\frac{2k}{\chi-\tomegaE}\right)-\oP(1)\log\left(\frac{2k}{\chi+\tomegaE}\right)}{k}+\frac{d_n}{k},
\end{equation}
where
\begin{equation}
    d_n= \int_{-1}^{+1} dv\, \frac{D(v) v^n-D(1)}{1-v}
\end{equation}
is a convergent integral. The logarithms are unambiguously defined for positive arguments, whereas for complex ones, one needs their phase from a prescription that descends directly from the $i\epsilon$ prescription in Eq.~\eqref{eq:def_In}. (Of course, Landau's $\epsilon$ is not the lepton asymmetry parameter.) By comparing the imaginary part originating from the $i\epsilon$ prescription for negative $(\chi-\tomegaE)/k$, we find that
\begin{equation}\label{eq:prescription_logarithm}
    \mathrm{log}\left[\frac{2k}{\chi-\tomegaE}\right]\to \mathrm{log}\left[-\frac{2k}{\chi-\tomegaE}\right]-i\pi\,\mathrm{sign}(k).
\end{equation}
This prescription fixes the interpretation of the logarithms in the complex plane; the same expression holds for antineutrinos with $\chi+\tomegaE$.

This expansion separates the dominant logarithmic singularity under the assumption $\tomegaE,\chi\ll k$. Inserting these expansions in the dispersion relation Eq.~\eqref{eq:dispersion}, we obtain
\begin{eqnarray}\label{eq:dispersion_logarithmic_finite_k}
    &&\left[P(1)\log\left(\frac{2k}{\chi-\tomegaE}\right)-\oP(1)\log\left(\frac{2k}{\chi+\tomegaE}\right)\right]\bigl(d_0+d_2-2d_1\bigr)
    \nonumber\\[1.5ex]
    &&\kern15em{}+d_0 d_2-d_1^2+k(d_0-d_2)-k^2=0.
\end{eqnarray}
Notice that $d_0+d_2-2d_1=D_0-D_1$, which for an uncrossed distribution is always positive, assuming $D(v)>0$. This form of the dispersion relation is generally true very close to the positive half of the light cone.

\subsection[Infinitely large \texorpdfstring{$k$}{}]{Infinitely large \texorpdfstring{\boldmath$k$}{}}

To proceed, we separate the further discussion in different regimes of wavenumbers and begin with the limit of large $k$, where Eq.~\eqref{eq:dispersion_logarithmic_finite_k} becomes
\begin{equation}\label{eq:dominant_logarithmic_dispersion}
    \left(\frac{\chi-\tomegaE}{2k}\right)^{P(1)}\left(\frac{\chi+\tomegaE}{2k}\right)^{-\oP(1)}=\exp\left[-\frac{k^2}{D_0-D_1}\right].
\end{equation}
In the fast flavor limit, where $\tomegaE=0$, the solution is
\begin{equation}\label{eq:asymptotic_light_cone}
    \chi=2k \exp\left[-\frac{k^2}{D(1)(D_0-D_1)}\right].
\end{equation}
Therefore, in the fast limit there are always two branches of purely real modes, neither damped nor growing, that get asymptotically close to being luminal. 

What happens now if $\tomegaE$ is very small but non-zero? Let us first take the case of normal ordering, with $\tomegaE>0$. In principle, we might expect two different solutions, with $\chi=\pm\tomegaE$ for asymptotically large $k$. However, the choice $\chi\simeq -\tomegaE$ does not actually lead to a solution, since the left-hand side of Eq.~\eqref{eq:dominant_logarithmic_dispersion} diverges while the right-hand side tends to 0. So there is only one possible solution with $\chi=\tomegaE+\delta$, with $\delta\ll \tomegaE$. We can now find the value of $\delta$ perturbatively. Let us first take $k>0$, which gives
\begin{equation}
    \delta=2k\exp\left[-\frac{k^2}{P(1)(D_0-D_1)}\right]\left(\frac{\tomegaE}{k}\right)^{\oP(1)/P(1)}.
\end{equation}
This mode is purely real and therefore stable. Notice that this mode lies \textit{above} the resonance cone of neutrinos, and therefore also of antineutrinos; it is effectively not resonating with any particle, and therefore is expected to be stable.

If $k<0$, the solution must be modified, because $(\chi+\tomegaE)/2k\simeq \tomegaE/k<0$, so that $\log(k/\tomegaE)\to \log(-k/\tomegaE)+i\pi$. This means that the value of $\delta$ is modified to
\begin{equation}\label{eq:chi_real}
    \delta=2k\exp\left[-\frac{k^2}{P(1)(D_0-D_1)}\right]\left(-\frac{\tomegaE}{k}\right)^{\oP(1)/P(1)} e^{-i\pi\frac{\oP(1)}{P(1)}}.
\end{equation}
Crucially, the frequency has developed an imaginary part
\begin{equation}
    \mathrm{Im}(\omega)=-2k\exp\left[-\frac{k^2}{P(1)(D_0-D_1)}\right]\left(-\frac{\tomegaE}{k}\right)^{\oP(1)/P(1)} \sin\left(\pi\frac{\oP(1)}{P(1)}\right).
\end{equation}
This imaginary part, determining the growth or damping of the mode, is entirely resonant, and comes from the neutrinos and antineutrinos moving along the light cone. If $\oP(1)=P(1)$, the imaginary part vanishes; the wave interacts in the same way with neutrinos and antineutrinos, so the net growth rate vanishes. We find that if $P(1)>\oP(1)$ the mode grows, while if $P(1)<\oP(1)$ the imaginary part becomes negative and the mode is damped. In turn, the real part
\begin{equation}
    \mathrm{Re}(\omega)=k+\tomegaE+2k\exp\left[-\frac{k^2}{P(1)(D_0-D_1)}\right]\left(-\frac{\tomegaE}{k}\right)^{\oP(1)/P(1)} \cos\left(\pi\frac{\oP(1)}{P(1)}\right).
\end{equation}
This is \textit{inside} the resonance cone both for neutrinos and for antineutrinos, so it resonates with both species.

Let us now turn to the case of inverted ordering ($\tomegaE<0$). In this case, $\chi\simeq\tomegaE=-|\tomegaE|$ is still an asymptotic solution in the limit of $k\to \infty$, but now the cases $k>0$ and $k<0$ are switched. We write $\chi=\tomegaE+\delta$ and seek solutions with $\delta\ll\tomegaE$. For negative $k$, the argument of the logarithm $2k/(\chi+\tomegaE)\simeq k/\tomegaE>0$ so the solution is purely real and coincides with Eq.~\eqref{eq:chi_real}. For positive $k$, the argument of $\log(k/\tomegaE)$ becomes negative again, but since now $k$ is positive, the replacement rule is $\log(k/\tomegaE)\to \log(-k/\tomegaE)+i\pi$ so that
\begin{equation}
    \delta=2k\exp\left[-\frac{k^2}{P(1)(D_0-D_1)}\right]\left(-\frac{\tomegaE}{k}\right)^{\oP(1)/P(1)} e^{i\pi\frac{\oP(1)}{P(1)}};
\end{equation}
for the imaginary part we find
\begin{equation}
    \mathrm{Im}(\omega)=2k\exp\left[-\frac{k^2}{P(1)(D_0-D_1)}\right]\left(-\frac{\tomegaE}{k}\right)^{\oP(1)/P(1)} \sin\left(\pi\frac{\oP(1)}{P(1)}\right).
\end{equation}
So also for this mode we find that for $P(1)>\oP(1)>0$ there is an instability, that is sourced by the resonance with the antineutrinos close to the light cone. Notice that the real part of $\mathrm{Re}(\delta)<0$, so that these modes are within the resonance cone of antineutrinos, but outside the resonance cone of neutrinos. An analogous reasoning can be made for the other side of the resonance cone, with $\omega\simeq -k$; in this case, one finds similar conclusions with inverted trends for the orderings, with the normal ordering showing an instability for $k>0$ and the inverted ordering for $k<0$. For these modes, the role of $P(1)$ and $\oP(1)$ is inverted, so that the growth rate is proportional to $\sin\bigl[\pi P(1)/\oP(1)\bigr]$.

\subsection{Boundary of unstable regions}

We have shown that at infinitely large $k$, unstable modes with a phase velocity very close to the speed of light exist, with a growth rate exponentially decreasing with $k$. As $|k|$ is lowered, the growth rate increases, but at some point it will decrease back to zero, marking the end of the unstable range of wavenumbers. The value of $k$ at the edge of the instability region provides an estimate for the minimum wavenumber that is unstable. To obtain it, we make the ansatz, confirmed by the result and motivated by numerical examples, that the unstable modes terminate with a superluminal velocity. In this case, we have shown elsewhere~\cite{Fiorillo:2024uki} that if the dispersion relation has the form $\Phi(\omega, k)=0$, the value of $k$ marking the edge of an unstable region is determined by the simultaneous condition $\partial \Phi(\omega,k)/\partial \omega=0$. In this case, using Eq.~\eqref{eq:dispersion_logarithmic_finite_k}, we immediately see that the frequency at which this happens is
\begin{equation}\label{eq:chi_threshold}
    \chi_{\rm thr}=-\tomegaE \frac{P(1)+\overline{P}(1)}{P(1)-\oP(1)}=-\tomegaE\frac{S(1)}{D(1)};
\end{equation}
notice that this is of order $\omega_{\rm thr}\sim \tomegaE/\epsilon$. Therefore, the wavenumber is now determined by the condition
\begin{eqnarray}
   \kern-1.5em P(1)\log\left(-\frac{k D(1)}{P(1)\tomegaE}\right)-\oP(1)\log\left(-\frac{k D(1)}{\oP(1)\tomegaE}\right)=
    \frac{k^2-k(d_0-d_2)+d_1^2-d_0 d_2}{D_0-D_1}.
\end{eqnarray}
Notice that for the unstable branches identified above, the arguments of the logarithms are always positive, as they should for a superluminal mode: for normal ordering ($\tomegaE>0$) and $P(1)>0$ and $\oP(1)>0$, the unstable modes are at $k<0$, while for inverted ordering ($\tomegaE<0$) they are at $k>0$.

This equation provides implicitly the value of $k$ at which the branch of unstable modes disappears. We are not interested in the precise value, but we will prove that for $\epsilon\ll 1$, this value is of the order of $k\sim \mu\epsilon$. This is an important physical statement: the typical wavelengths becoming unstable are very short, of the same order of magnitude as those of fast-unstable modes. To show this, we note that in the limit $\epsilon\ll 1$ we may write $P(1)\simeq \oP(1)\simeq S(1)/2$, and introducing $\kappa=-k D(1)/S(1) \tomegaE>0$ we find
\begin{equation}\label{eq:finding_threshold}
    \log(2\kappa)-\frac{S(1)^2 \kappa^2 \tomegaE^2+D(1) (D_0-D_2) S(1)\kappa \tomegaE+D(1)^2(D_1^2-D_0 D_2)}{D(1)^3 (D_0-D_1)}=0.
\end{equation}
The function on the left-hand side is negative for $\kappa\to 0$ and passes through zero near $\kappa\simeq 1/2$, where the logarithm vanishes and the remaining terms are of order $\tomegaE^2/\mu^2\epsilon^4\ll 1$. However, this solution is not acceptable, since it gives a wavenumber in order of magnitude $k\sim \tomegaE/\epsilon$, but $\chi$ is also $\chi\sim \tomegaE/\epsilon$ from Eq.~\eqref{eq:chi_threshold}, violating our original assumption that $\chi\ll k$. However, at larger values of $\kappa$, the function on the left-hand side of Eq.~\eqref{eq:finding_threshold} is certain to vanish again, since at $\kappa\to \infty$ it becomes again negative due to the term proportional to $\kappa^2$. Thus, the function, after passing through zero at $\kappa\simeq 1/2$, will have a maximum and then decrease, passing through zero again. The position of this second zero can be found in order of magnitude by finding the value of $\kappa$ at which the function has a maximum, since the zero will be at comparable values of $\kappa$. By the condition of vanishing derivative, we find that the maximum of the function is at a value of $\kappa$ such that
\begin{equation}
    D(1)^3 (D_0-D_1)+D(1)(D_2-D_0)S(1)\kappa \tomegaE-2S(1)^2 \kappa^2 \tomegaE^2=0.
\end{equation}
From the structure of this equation, we see that the solution will be at values of $\kappa\simeq \mu\epsilon^2/\tomegaE\gg1$, in turn implying $k\sim \epsilon$. Therefore, the unstable modes will terminate, in the limit $\epsilon \ll 1$ and $\tomegaE\ll \mu\epsilon^2$, at very large wavenumbers of order $k\sim \mu\epsilon$.

\subsection{Luminal unstable wavenumbers}

As we have seen, the unstable modes pass from being superluminal to subluminal. We can therefore guess that the maximum growth rate will be when the phase velocity is very close to luminal. This motivates us to look into solutions with $\chi$ purely imaginary, so we write $\chi=i \chiI$. We now assume, confirmed by the result, that for $\epsilon\ll 1$ we have $\chiI\gg \omegaE$. Therefore, in Eq.~\eqref{eq:dispersion_logarithmic_finite_k}, we can expand the logarithmic term to find
\begin{equation}
    D(1)\log\left[\frac{2k}{i\chiI}\right]+\frac{S(1)\tomegaE}{i\chiI}+f(k)=0,
\end{equation}
with
\begin{equation}
    f(k)=\frac{d_0d_2-d_1^2+k(d_0-d_2)-k^2}{D_0-D_1}.
\end{equation}
From the imaginary part of this equation, we can now extract the growth rate; the sign of the imaginary part of the logarithm is fixed by the prescription in Eq.~\eqref{eq:prescription_logarithm}, so that
\begin{equation}
    \chiI=-\frac{2\tomegaE S(1) \mathrm{sign}(k)}{\pi D(1)}.
\end{equation}
We find correctly that for $k>0$, the modes are unstable in inverted ordering ($\tomegaE<0$), whereas for $k<0$, the modes are unstable in normal ordering ($\tomegaE>0$). More importantly, we find that the typical growth rate is of the order of magnitude $\chiI\sim \tomegaE/\epsilon\gg \tomegaE$, as we had initially assumed. The growth rate for these modes vanishes when $P(1)+\oP(1)=0$.

\subsection{Summary}

The main novelty found in this section is that in the limit of very small vacuum frequency, $|\tomegaE|\ll \mu\epsilon^2$, there exist slow instabilities. Their physical nature is however completely different from the usual ones. These instabilities appear at rather large wavenumbers (small spatial scales) of the order of $k\sim \mu \epsilon$, comparable with the length scales of fast instabilities. The growth rates can reach a maximum of the order of $\mathrm{Im}(\omega)\sim \tomegaE/\epsilon$; thus, for distributions with a nearly equal amount of neutrinos and antineutrinos, they can be much faster than the vacuum oscillation frequency, yet not as fast as the often-quoted $\sqrt{\tomegaE\mu}$ scaling. These unstable modes survive up to arbitrarily large wavenumbers, but with exponentially suppressed growth rates. In the next section, we will put together the pieces proved in Sects.~\ref{sec:non-resonant} and~\ref{sec:resonant} and show how these modes appear in a specific example.

We have not discussed the axial-breaking modes. One can perform an analogous expansion near the light cone, but the results are less illuminating because these modes cannot exist close to the light cone at arbitrarily large $k$. The reason is seen from Eq.~\eqref{eq:dispersion_transverse}; at very large $k$, the integrals $I_0-I_2$ converge to $0$, because the integrand function does not have any divergence when $\omega$ is very close to the light cone given the factor $(1-v^2)$ in the numerator. Numerically we will still find that modes close to the light cone generally exist at finite $k$, and that for normal ordering they do become unstable, as expected since they provide the initial unstable modes which, at $\omega_E\gg \mu \epsilon^2$, turn into the non-resonant unstable dipole modes we have found in Sec.~\ref{sec:non-resonant}.

\section{Pattern of slow instabilities}
\label{sec:numerical}

In this section, we summarize our conclusions on the extreme regimes $|\tomegaE|\ll \mu \epsilon^2$ and $|\tomegaE|\gg \mu \epsilon^2$, and extend them to the intermediate regime, on the basis of an explicit numerical example for a benchmark angular distribution. It is uncrossed so that the nature of the slow instability is not contaminated by any interplay with the fast one.

\subsection{Setup of the benchmark system}

\begin{figure}
    \includegraphics[width=\textwidth]{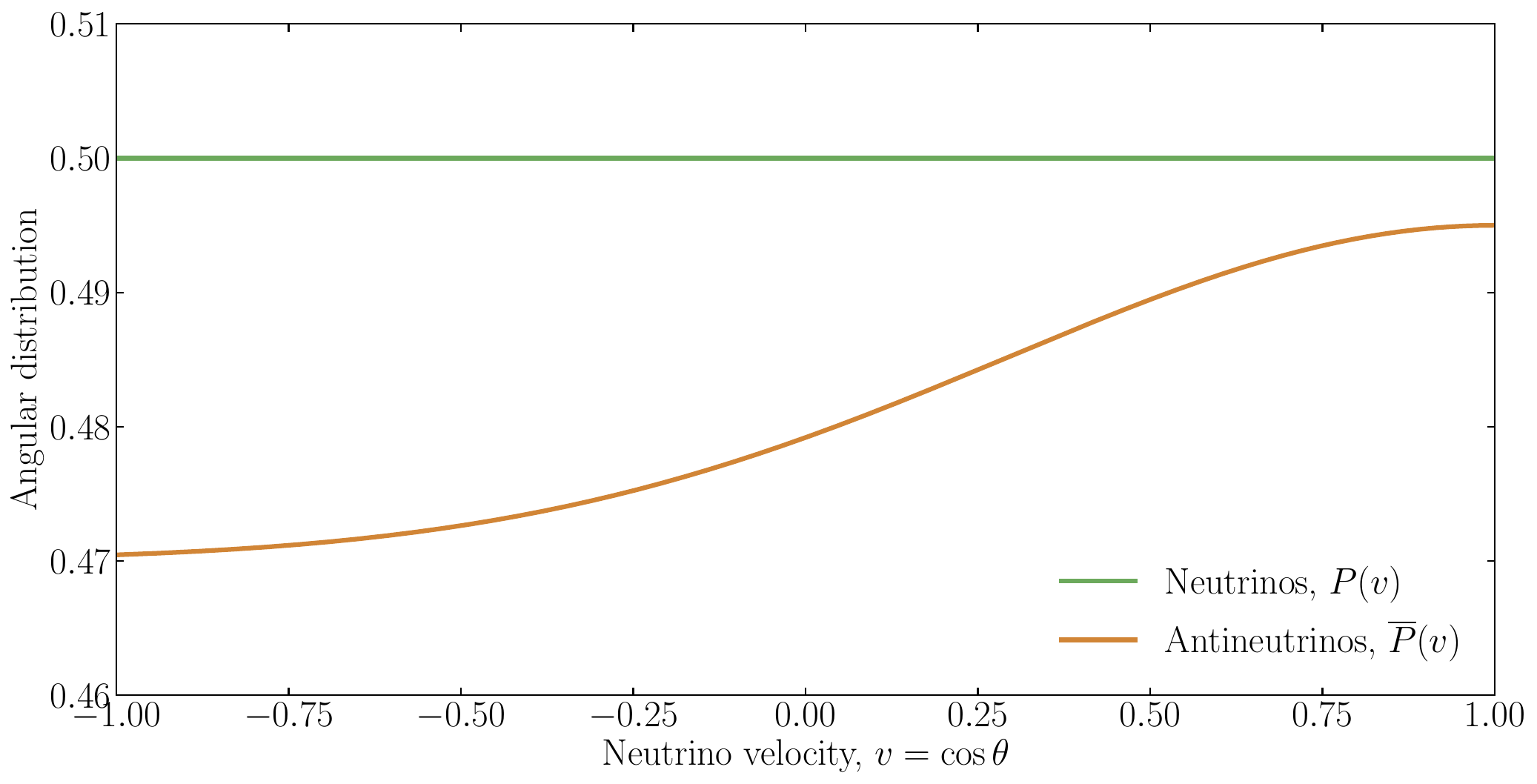}
    \caption{The benchmark angular distribution for neutrinos and antineutrinos
    defined in Eq.~\eqref{eq:benchmark} and used to exemplify the pattern of collective modes. \label{fig:benchmark}}
\end{figure}

The functional form of our uncrossed benchmark distribution is inspired by examples used in the previous literature \cite{Yi:2019hrp, Padilla-Gay:2021haz, Shalgar:2024heq} and shown in Fig.~\ref{fig:benchmark}. Specifically, for the neutrinos, we use an isotropic distribution, whereas for the antineutrinos, we use a small deviation from isotropy, overall of the form
\begin{equation}\label{eq:benchmark}
    P(v)=0.5
    \quad\hbox{and}\quad
    \oP(v)=0.47+0.025\exp\left[-(1-v)^2\right].
\end{equation}
If there is no crossing, why not use a completely isotropic distribution? The motivation for our choice is that on the one hand, an isotropic distribution does not have simpler analytical properties for the collective modes, but on the other hand, it is too specialized. The singular character of the isotropic distribution is found even in the fast limit ($\tomegaE=0$), where no Landau-damped modes exist\footnote{We emphasize that this is true only for the longitudinal modes; the axial-breaking modes exhibit a Landau-damped branch even for an isotropic distribution.}. In the fast case, the latter can only appear or disappear on the light cone, namely for $\omega/|\bk|=\pm 1$, an argument which actually applies for $\bk$ modes along any direction, not just along the axis of symmetry. But for an isotropic distribution, the luminal sphere $\omega/|\bk|=\pm 1$ does not have any privileged direction, and therefore Landau-damped modes cannot originate anywhere on it. The absence of such modes is special for a perfectly isotropic distribution; even a small amount of anisotropy introduces Landau-damped modes which greatly simplify the analytical properties of the collective modes. For our example, the difference between neutrino and antineutrino distributions is on the percent level of their sum (actually $\epsilon \simeq 0.019$), a value that we will use to verify the scaling laws predicted in the previous sections. 

In the following sections, we will show the collective modes that are found for different $\tomegaE$ values by solving the dispersion relation in Eq.~\eqref{eq:dispersion}. Before presenting the results, we stress that finding these solutions is numerically far from trivial. Using a dense set of discrete velocity modes rather than a continuous distribution is not a good strategy in the context of slow instabilities. Even for fast instabilities, one loses the Landau-damped modes, which are natural continuations of the unstable modes but only appear in the continuous limit. More importantly, in the absence of slow instabilities, the majority of modes are Case-Van-Kampen modes that lie below the light cone~\cite{Fiorillo:2023mze, Fiorillo:2024bzm}. If the system has slow instabilities, these modes acquire imaginary parts, seemingly becoming unstable, providing spurious instabilities that disappear as the number of modes grows. 

Therefore, we solve the transcendental equation Eq.~\eqref{eq:dispersion} directly. While the required numerical integration fundamentally involves a discretization, one does not solve a polynomial equation and spurious modes do not show up. On the other hand, for a given value of $k$ it may admit many different true solutions of complex $\omega$, so it is not easy to ensure that all of them are found. In other words, here one may miss existing modes instead of producing spurious ones. We use automatic numerical algorithms of root finding, that require a starting point to seek the zero of the dispersion relation. To this end, we first find the solutions for $k=0$, where the dispersion relation is algebraic and has exactly four solutions (that become two degenerate solutions in the fast limit of $\tomegaE=0$). We then increase or decrease $k$ slowly, using for each value of $k$ the previous solution as a starting point. However, this procedure does not find those branches that disappear before reaching $k=0$. Such branches generally exist \cite{Fiorillo:2024bzm} because Landau-damped modes can disappear abruptly when they touch one of the branch roots $\mathrm{Re}(\omega)=k\pm \tomegaE$. In these cases, we explicitly search for solutions with negative imaginary part with the numerical root-finding algorithm, but it is not necessarily guaranteed that one finds all solutions.

\subsection[Fast case (\texorpdfstring{$\tomegaE=0$}{})]{Fast case (\texorpdfstring{\boldmath$\tomegaE=0$}{})}

\begin{figure}
    \includegraphics[width=\textwidth]{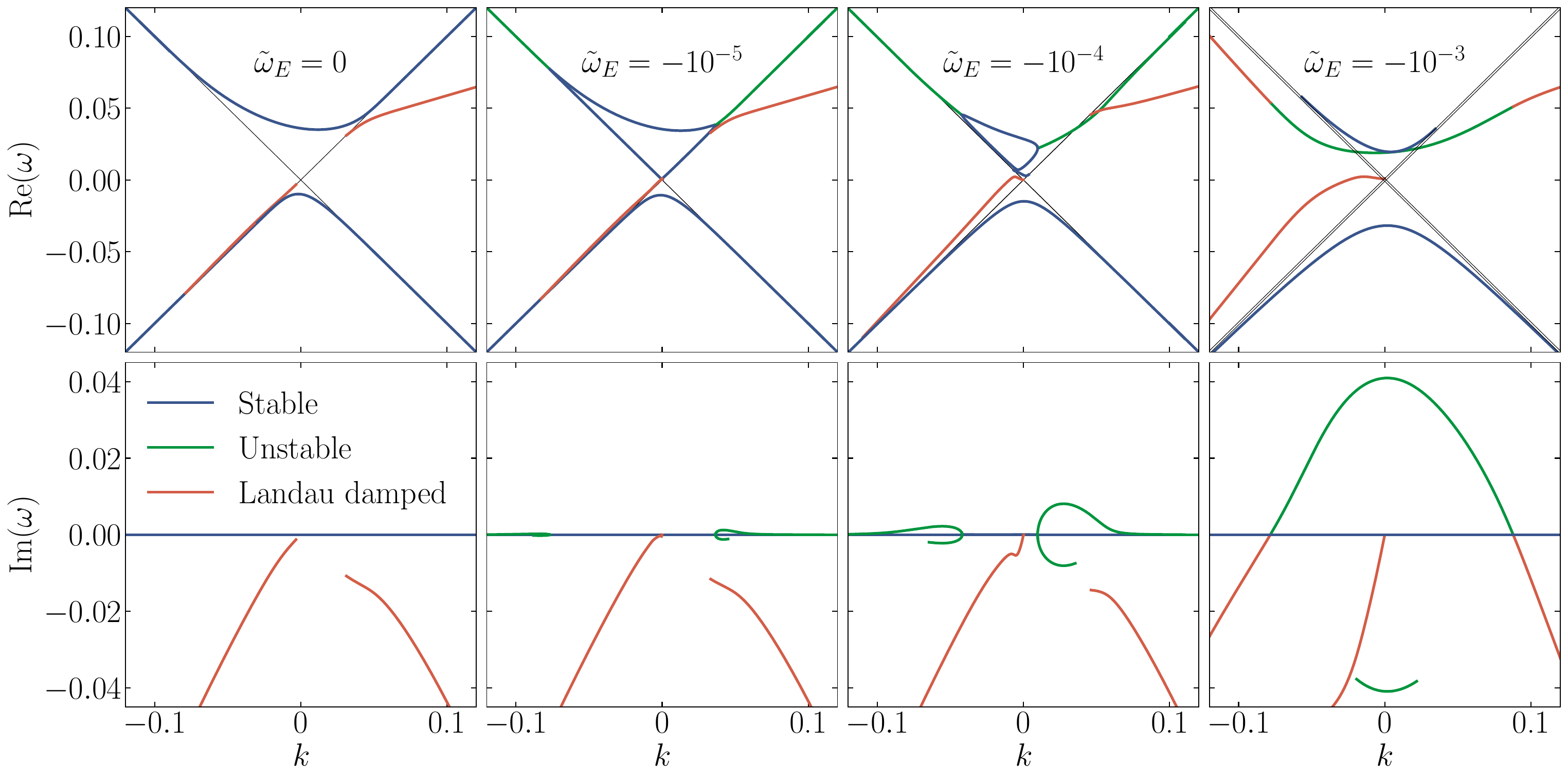}
    \includegraphics[width=\textwidth]{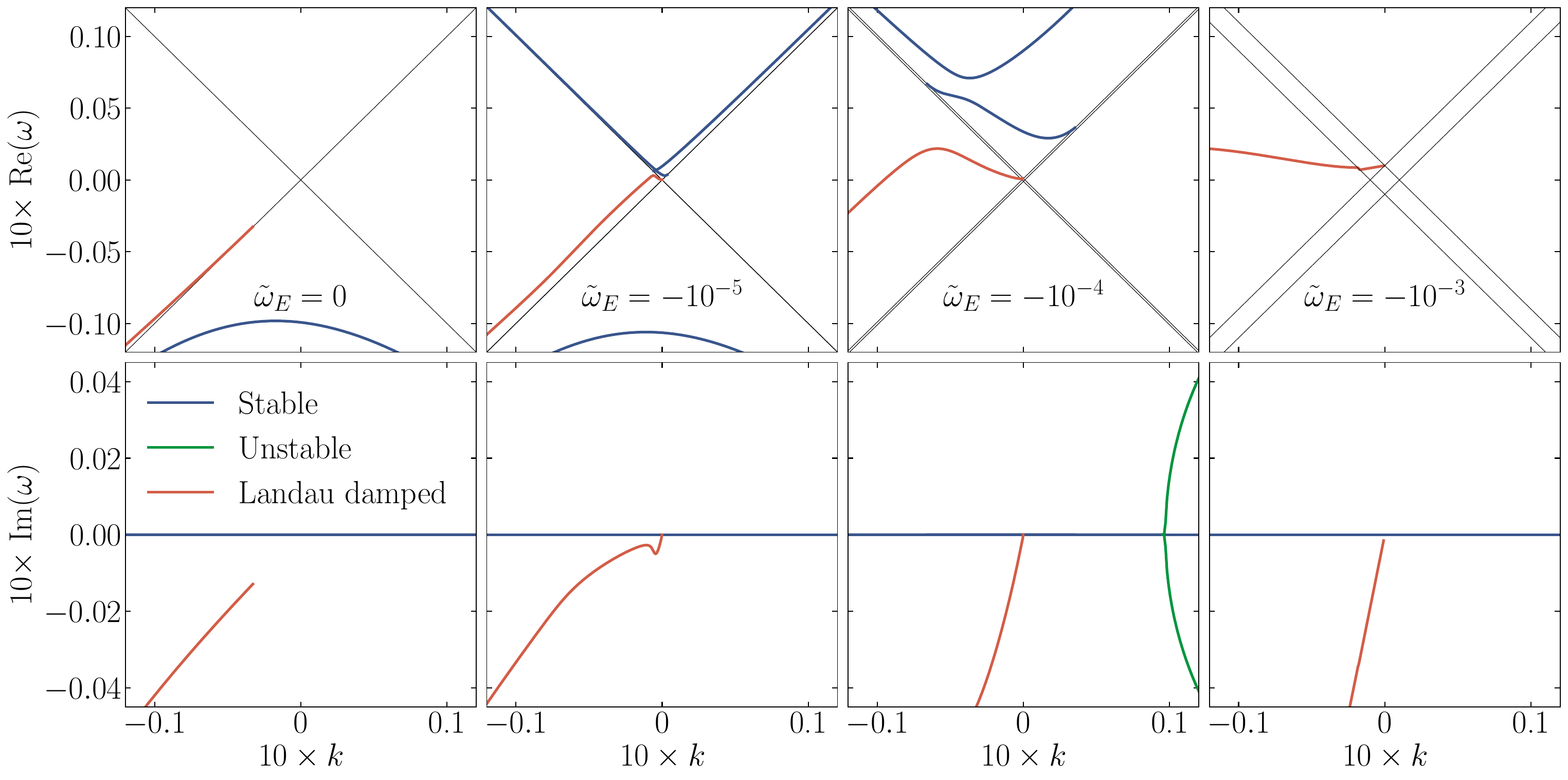}
    \vskip-6pt
    \caption{\label{fig:inverted_ordering} Dispersion relation for inverted mass ordering following from Eq.~\eqref{eq:dispersion} and the angular spectrum of Eq.~\eqref{eq:benchmark}. The vacuum oscillation frequency $\tomegaE<0$ is indicated in the panels. We distinguish real branches (blue), unstable ones (green), and Landau-damped ones (red). With the same color as the unstable branches, we also show the damped superluminal modes that are complex conjugate to an unstable mode. We also show as black diagonal lines the resonance cones discussed in Sec.~\ref{sec:resonancecones} and defined by the condition $\omega=\pm k\pm\tomegaE$.}
\end{figure}

Turning now to our explicit numerical results, we begin with the case of vanishing vacuum frequency ($\tomegaE=0$). The panels in the left column of Fig.~\ref{fig:inverted_ordering} show both $\mathrm{Re}(\omega)$ and $\mathrm{Im}(\omega)$ as a function of $k$. In addition to the full structure over a wide range of $k$ (upper panels) we also show a zoomed-in version focusing on a smaller range of $k$ and $\omega$ close to the crossings of the light cone (lower panels).

Qualitatively, the fast case ($\tomegaE=0$) can be easily understood using the concepts introduced in Refs.~\cite{Fiorillo:2024bzm, Fiorillo:2024uki}. There are two branches of purely stable modes that asymptotically approach the light cone for $\mathrm{Re}(\omega)\to \pm \infty$. We have already proved the existence of these modes in Eq.~\eqref{eq:asymptotic_light_cone}. In addition, we identify two branches of Landau-damped modes that originate on the luminal sphere. The left branch ($k<0$) covers only a small range of wavenumbers, being born and dying very rapidly on the light cone. The right branch leaves the light cone very rapidly, and does not turn back towards it. We have verified that this trend proceeds up to very large $k$. 

The reason for this behavior, which is so asymmetrical between the left and right branch of Landau-damped modes, comes ultimately from the exponential term in the distribution $P(v)$; when $\mathrm{Im}(\omega)<0$, the integrals $I_n$ receive a contribution from the part of the integration path surrounding the pole in the lower half-plane $u=\omega/k$. Since $\oP(v)\propto e^{-(1-v)^2}$, this contribution $\oP(u)$ becomes exponentially large as $\mathrm{Im}(\omega)$ grows. Thus for $k>0$, even when $k$ becomes very large, the functions $I_n$ can remain finite because their numerator becomes increasingly large, leading to the existence of Landau-damped modes up to very large $k$. The fact that Landau-damped modes can depend so sensitively on the specific functional form of $P(v)$ and $\oP(v)$, and on its analytic continuation for complex $v$, may at first seem unphysical. On the other hand, the modes that depend so sensitively on the analytic continuation of the function far from its real argument are essentially the modes with a very large damping rate, which therefore become irrelevant over timescales so short as to be inessential. The Landau-damped modes that are physically most relevant are those with a small value of $\mathrm{Im}(\omega)$, and they are determined by the functions $P(v)$ and $\oP(v)$ evaluated close to the real axis.

\subsection[Inverted ordering (\texorpdfstring{$\tomegaE<0$}{})]{Inverted ordering (\texorpdfstring{\boldmath$\tomegaE<0$}{})}

Figure~\ref{fig:inverted_ordering} also shows the changes in the structure of the dispersion relation as we increase $|\tomegaE|$ in the regime of inverted ordering ($\tomegaE<0$). As soon as a small nonvanishing $\tomegaE=-10^{-5}$ is introduced, the branch in the upper part of the light cone effectively splits into pairs of modes. At large $|k|$, in the upper light cone, the unstable modes that we had anticipated in Sec.~\ref{sec:resonant} appear. In this region, the modes are subluminal and originate from the resonant wave-particle interaction with collinear particles. As expected, their typical wavenumbers are of order $|k|\sim \mu \epsilon$, and as their wavenumber decreases they transition into two branches of real modes that remain very close to the light cone.  We recall here that unstable superluminal modes always come with a complex conjugate, which is visible in the panels showing $\mathrm{Im}(\omega)$ -- we show it in the same color as the unstable modes -- but the latter disappears when the mode passes through the light cone. This behavior is as expected, since the dispersion relation has a branch cut for $\mathrm{Re}(\omega)=k\pm \tomegaE$ and $\mathrm{Im}(\omega)<0$, so damped superluminal modes can disappear into the branch cut. We should also stress that the two unstable bumps on the two sides are asymmetrical because they originate from the resonant interaction with collinear neutrinos, which are moving with $v\simeq -1$ for $k<0$ and with $v\simeq +1$ for $k>0$, and the amount of neutrinos along $v=\pm 1$ is different. Thus, if the angular distribution were ``flipped,'' with $P(v)\to P(-v)$ and $\overline{P}(v)\to \overline{P}(-v)$, the two bumps would exchange place.

At very small $k$, a third branch of real modes appears, which is only visible in the zoomed version, because it disappears at extremely small $k$. While this real branch is not fundamental, since it does not develop any instability, it is still relevant since it ensures that at $k=0$ we have, as expected, four solutions, corresponding to the four solutions of the algebraic dispersion relation. Meanwhile, the Landau-damped modes and the branch of real modes in the lower part of the light cone are only weakly affected by $|\tomegaE|$, as we had also predicted in Sec.~\ref{sec:resonant}.

As we increase $|\tomegaE|$, the unstable modes extend to lower $|k|$, while the stable branch of real modes between them shrinks. The growth rate of the unstable eigenmodes also visibly increases, essentially linearly, remaining of the order of magnitude of $|\tomegaE|/\epsilon$, again as predicted; for $\tomegaE=-10^{-4}$, we find a growth rate of order $\mathrm{Im}(\omega)\sim 0.01$, consistent since $\epsilon\sim 0.01$. Notice that with this value of $\epsilon$ we also expect $|\tomegaE|\sim \mu \epsilon^2\sim 10^{-4}$ to lie at the transition of the emergence of non-resonant modes. Indeed, we find that for $\tomegaE=-10^{-3}$, the branches of unstable modes at large wavenumbers have finally merged into a single branch of unstable modes at all wavenumbers, including $k=0$. At this point, we recover the scaling $\mathrm{Im}(\omega)\sim \sqrt{|\tomegaE|\mu}$ well known from the case of the flavor pendulum. 

At larger wavenumbers, the unstable branch merges with the right Landau-damped branch, which effectively disappears. The other branch of Landau-damped modes, which originally lied close to the lower light cone, remains only perturbed by $\tomegaE$ but is not qualitatively affected; this is all consistent with our general finding that in the case of inverted ordering, $\tomegaE$ primarily affects the branches close to the upper light cone ($\omega>0$), but not close to the lower light cone ($\omega<0$).

\subsection[Normal ordering (\texorpdfstring{$\tomegaE>0$}{})]{Normal ordering (\texorpdfstring{\boldmath$\tomegaE>0$}{})}

\begin{figure}
    \includegraphics[width=\textwidth]{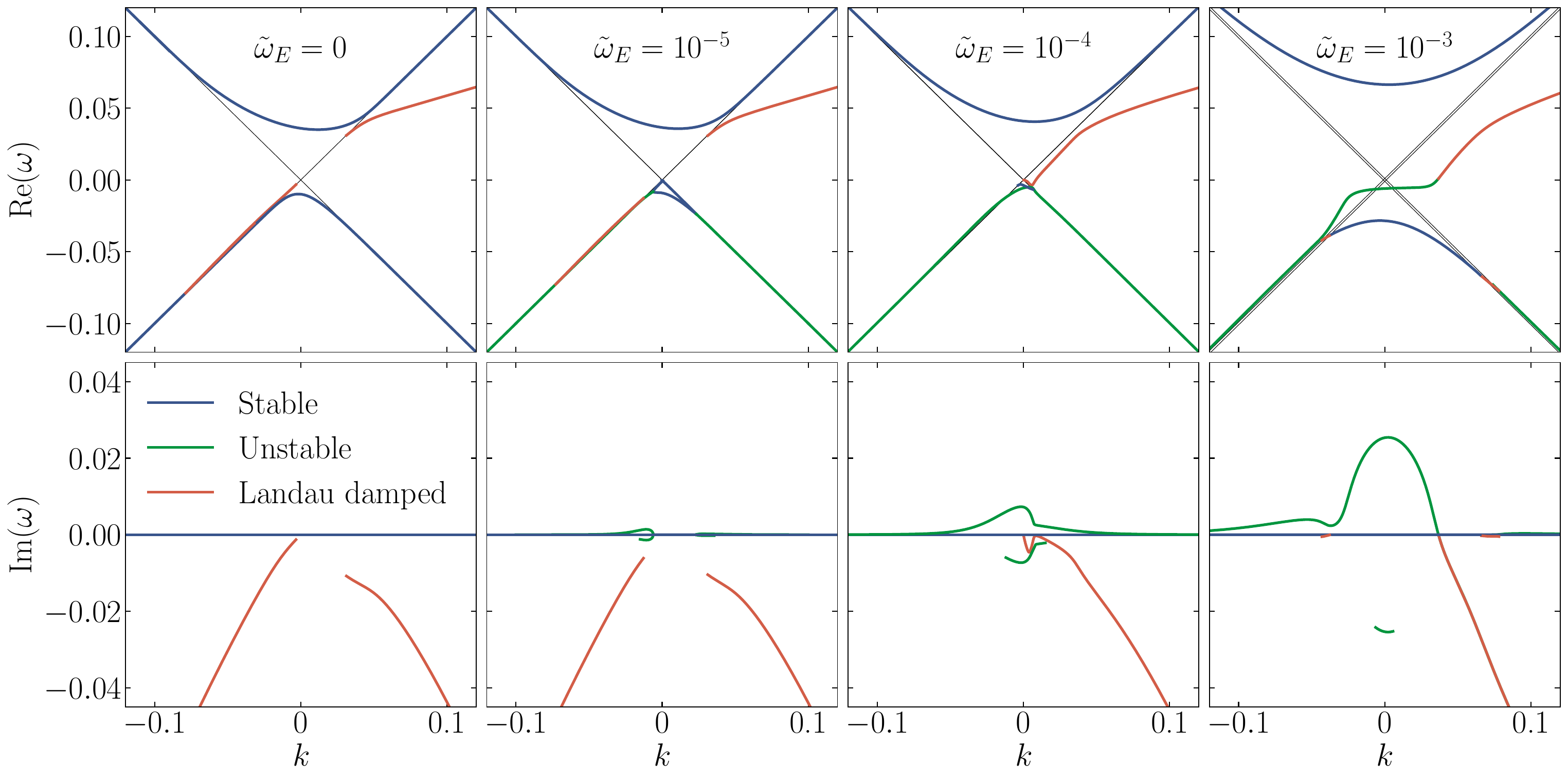}
    \includegraphics[width=\textwidth]{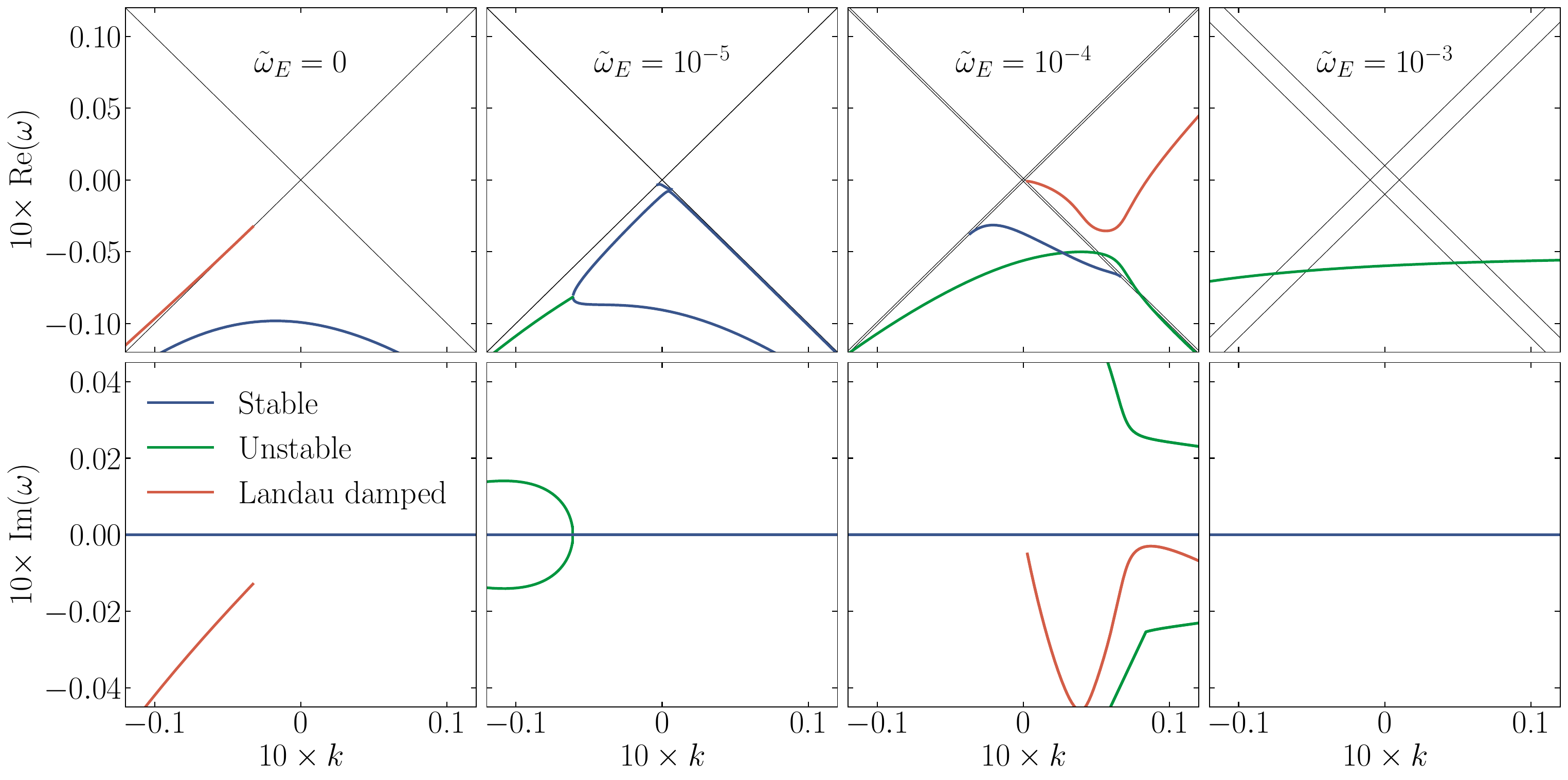}
    \caption{\label{fig:normal_ordering}Same as Fig.~\ref{fig:inverted_ordering} for normal ordering ($\tomegaE>0$).}
\end{figure}

Next we turn to normal ordering ($\tomegaE>0$) and show the analogous results for the dispersion relation in Fig.~\ref{fig:normal_ordering}. Similar to the previous case, as soon as $\tomegaE$ becomes nonzero, unstable branches appear, but this time around the lower light cone, as expected. Compared to inverted ordering, the unstable branches now reach very rapidly down to very small values of $|k|$, presumably because in the fast limit ($\tomegaE=0$) the lower real branch sticks much closer to the light cone than the upper one. As we see from the zoomed version, we still have three real branches for $\tomegaE=10^{-5}$ in the lower light cone, while the real branch in the upper light cone is only quantitatively, but not qualitatively, affected by $\tomegaE>0$.

Since the unstable modes reach much faster to low wavenumbers than for inverted ordering, it is not surprising that at $\tomegaE=10^{-4}$ the two unstable modes have already merged into a single branch; close to $k=0$, this branch is by definition non-resonantly unstable. The Landau-damped modes around the upper light cone are at this stage unaffected.

However, as $\tomegaE$ increases to $10^{-3}$, we find that in normal ordering the transition to the fully non-resonant regime, again happening around $\tomegaE\sim \mu \epsilon^2 \sim 10^{-4}$, is quite different. The unstable branch at $\tomegaE=10^{-3}$ here absorbs both of the previous Landau-damped branches, effectively passing from the lower to the upper light cone. The difference to inverted ordering, where instead the unstable branch remained close to the upper light cone, is therefore quite noticeable. An instability at very large positive $k$ still exists, but it does not belong to the same branch as the non-resonant unstable modes. Rather, it smoothly continues into the real band in the lower light cone, which previously (at $\tomegaE=10^{-5}$) existed only at very small wavenumbers.

The simple conclusion that the vacuum frequency in inverted ordering mostly affects the upper part of the light cone ($\omega>0$), while in normal ordering mostly affects the lower part ($\omega<0$), can be understood through various arguments. The analytical expansions we have performed in Sec.~\ref{sec:resonant} show this explicitly, and provide a straightforward approximation for the growth rates and regions of instability. On the other hand, a simple and less rigorous argument can be made. As we have seen in Secs.~\ref{sec:homogeneous-isotropic} and~\ref{sec:non-resonant}, in inverted ordering the mode that becomes unstable for a near-isotropic distribution in the non-resonant regime is the monopole mode. In the fast limit, this mode satisfies $\omega\simeq D_0>0$. Thus, since this mode must ultimately become unstable as we increase $\tomegaE$, we reasonably expect that the modes with $\omega>0$ are the ones qualitatively changed by the introduction of $\tomegaE$. On the contrary, for normal ordering, it is the dipole mode that will ultimately become unstable. In the fast limit, this mode satisfied $\omega\simeq -D_2<0$, and therefore we expect qualitative changes induced by the vacuum frequency primarily around the lower light cone $\omega<0$.

\subsection{Axial-breaking modes in normal ordering}

\begin{figure}
    \includegraphics[width=\textwidth]{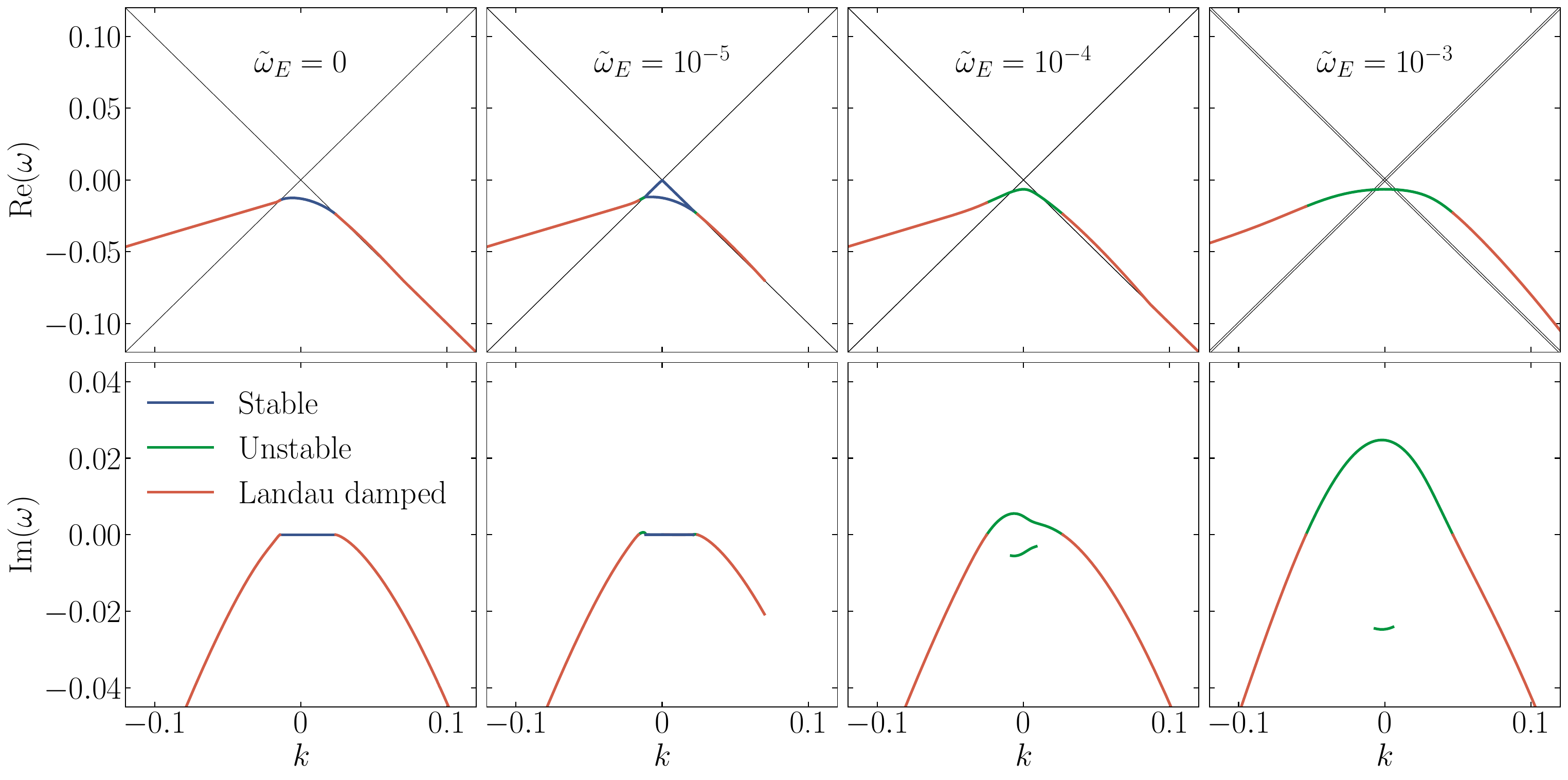}
    \vskip-6pt
    \caption{\label{fig:axial_breaking}Same as Fig.~\ref{fig:inverted_ordering} for axial-breaking modes in normal ordering ($\tomegaE>0$).}
\end{figure}

Finally, we turn to axial-breaking modes that must satisfy Eq.~\eqref{eq:dispersion_transverse} and show the dispersion relation in Fig.~\ref{fig:axial_breaking}. We consider only normal ordering, since for inverted ordering no unstable modes appear. This result is consistent with our earlier finding in Sec.~\ref{sec:non-resonant} that for $|\tomegaE|\gg \mu \epsilon^2$ the axial-breaking modes become unstable only for normal ordering, and this conclusion is found to pertain also for smaller values of $|\tomegaE|$.

In the fast case ($\tomegaE=0$), we correctly find that at very large wavenumbers there are no stable modes close to the light cone. Instead, there are two branches of Landau-damped modes close to the lower light cone, which smoothly connect with the Landau-damped modes on both sides. The reason that there can be a smooth connection between these modes across the light cone is easy to understand once we notice that the combination of integrals $I_0-I_2$, appearing in Eq.~\eqref{eq:dispersion_transverse}, does not have a discontinuity or an infinity at the light cone ($\omega=\pm k$), because the numerator of the integrand function contains the factor $1-v^2$ which vanishes on the light cone. Notice that in the fast case there is no branch close to the upper light cone. 

As we turn on the vacuum frequency $\tomegaE=10^{-5}$, the Landau-damped modes at large $k$ are not strongly affected, but close to the light cone, where they were previously transitioning to real modes, they now pass through a brief phase of instability. Once entering the superluminal regime, out of the light cone, the pairs of complex conjugate unstable modes split into two real branches which stick close to the lower light cone. As we increase $\tomegaE$ to $10^{-4}$, the two unstable branches on the two sides of the light cone merge, providing a single unified branch across the lower light cone. Finally, when $\tomegaE=10^{-3}$, the growth rate of the mode starts to increase in the typical $\mathrm{Im}(\omega)\sim \sqrt{|\tomegaE|\mu}$ fashion, effectively reproducing the non-resonant unstable dipole mode in normal ordering.

\section{Summary and discussion}
\label{sec:conclusions}

The phenomenology of collective unstable modes of flavor conversion has had a somewhat scattered historical development. The original discovery of unstable modes came from studies of neutrinos in the early universe with an assumed homogeneous and isotropic distribution and the instability was driven by the vacuum frequency splitting between neutrinos and antineutrinos, what today we would call the slow flavor pendulum. Later these ideas were mapped on the spatial evolution of SN neutrinos, assuming static solutions driven by a stationary source in the form of the bulb model of emission. At the time when the idea of purely static solutions was slowly recognized to be unsustainable, the relevance of fast dynamics, driven by crossed angle distributions, was finally acknowledged (although had been proposed much earlier) and took up the attention of our community to develop the theory and phenomenology of fast flavor conversion in the limit of vanishing neutrino masses. We here bridge the gap between the theory of fast and slow modes, i.e., apply the novel language of space-dependent temporal growth, inspired by plasma physics, to understand how the vacuum frequency can induce instabilities despite being so much smaller than the refractive energy scale. We show that many seemingly generic features of homogeneous and isotropic slow instabilities are actually special to large-scale modes. 

We have also stressed that the driving parameter $\tomegaE=(\delta m^2/2E)\cos\tV$, that we call the vacuum oscillation frequency, actually involves the projection by the vacuum mixing angle on the weak-interaction direction in flavor space. Even in the presence of large matter refraction, this projection involves the true vacuum mixing angle, not an effective in-medium mixing angle. We have not investigated the impact of matter on the slow-oscillation phenomenology that would manifest in the nonlinear regime.

Our first new insight is that the behavior of slow instabilities depends on a parameter whose relevance is usually under-emphasized, namely the ratio between lepton number and particle number, which we dub $\epsilon$. The impact of the vacuum frequency on the pattern of slow instabilities is completely different in the regimes of $|\tomegaE|\ll \mu \epsilon^2$ or $|\tomegaE|\gg \mu \epsilon^2$.  We have restricted our study to a few simplifying assumptions, including axisymmetric angular distributions with no angular crossing, allowing for a clean separation of the slow modes from the potential fast ones.

The first regime with very small vacuum frequencies $|\tomegaE|\ll \mu \epsilon^2$ is physically the regime of large neutrino density if the vacuum oscillation frequency is taken as a fixed parameter. In this regime, large-scale modes, and notably the slow flavor pendulum, are stable even in the inverted position, what has been dubbed the sleeping-top regime. Instabilities here depend on inhomogeneous solutions that break the initial homogeneity. The growth rates $\mathrm{Im}(\omega)$ of such modes are of the order of $|\tomegaE|/\epsilon$, but only for very small-scale modes, with a length scale comparable to that of fast unstable modes $(\mu \epsilon)^{-1}$. The main feature we have highlighted is the resonant nature of these modes, since they come primarily from the wave-particle interaction with neutrinos moving close to the axis of symmetry. 

However, this conclusion descends primarily from our consideration of modes directed only along the axis of symmetry. Modes in other directions would resonate with neutrinos along these directions; this is particularly evident for an isotropic distribution, which is axisymmetric in any direction. Thus there are small-scale modes resonating with neutrinos along any direction; their main feature is a phase velocity close to the speed of light. This version of slow unstable modes has not been highlighted before and is phenomenologically particularly relevant. It allows, in principle, for local relaxation even for large neutrino densities as in a SN core, over a timescale of order $\epsilon/\tomegaE$, which is generally short compared to the collisional one. Notice also that, while our benchmark example was for a near-isotropic distribution, our analytical treatment shows that these distributions are resonant, and therefore depend on local properties of the angular distribution for neutrinos moving in phase with the mode. Thus, even if the distribution is not near-isotropic, the existence and general properties of these modes as inferred in Sec.~\ref{sec:resonant} remain similar. Therefore, phenomenologically these slow instabilities can be relevant both in deeper SN regions where the angular distribution is nearly isotropic or at larger radii where it is more forward peaked. Since $\tomegaE$ is much smaller than the collision rate, these instabilities, despite being very slow compared to the fast ones, might still allow for a collisionless relaxation.

In the opposite regime with $|\tomegaE|\gg \mu \epsilon^2$ (but still $|\tomegaE|\ll \mu$), the system exhibits instabilities of the same nature as the well-studied slow flavor pendulum. These modes can be unstable on all length scales down to values of order $(\mu\epsilon)^{-1}$, thus also including large-scale modes. Their nature is non-resonant, at least in the large-scale region where their growth rate is largest, meaning that they arise from wave-particle interactions with the entire neutrino angular range. This is why they were the first to be discovered -- they appear easily also in cases of homogeneous and isotropic setups. The typical growth rate for these instabilities corresponds to the often-cited scaling $\mathrm{Im}(\omega)\sim \sqrt{|\tomegaE|\mu}$. This scaling motivates the terminology of {\em slow\/} modes and suggests a lesser relevance than fast modes driven purely by refraction. However, this scaling applies \textit{only} in the regime $|\tomegaE|\gg \mu \epsilon^2$ and therefore $\mathrm{Im}(\omega)$ is actually much \textit{larger} than the growth rate of fast instabilities, which usually is of the order of $\mu \epsilon$. Thus, in the regime in which the $\sqrt{|\tomegaE|\mu}$ scaling applies, slow modes are more rapidly growing than fast modes. 

The catch to this argument is that, unless $\epsilon$ is very small, in a SN core we usually expect to be in the first regime ($|\tomegaE|\ll \mu \epsilon^2$), where slow modes are even slower than usually envisioned, with $\mathrm{Im}(\omega)\sim |\tomegaE|/\epsilon$. The difference between the regimes of resonant and non-resonant instabilities may have deep consequences on the final outcome, although we have here not pursued the question of non-linear evolution. However, following the general quasi-linear picture of relaxation \cite{Fiorillo:2024qbl}, in addition to producing turbulent flavor fluctuations, the instability affects primarily the spatially averaged angular distribution along the directions that interact with the growing waves. In the non-resonant case, the entire angular distribution should be affected, and so to remove the original cause of instability, equipartition along any direction is generally expected over timescales $1/\sqrt{\tomegaE\mu}$. This is indeed what was found, e.g., in a homogeneous setup with spontaneous breaking of isotropy \cite{Raffelt:2007yz}, where the outcome was an average equipartition across the entire angular distribution, together with fluctuations which, in the framework of Ref.~\cite{Fiorillo:2024qbl}, we can interpret as turbulent in nature. This conclusion follows even if only the modes directed along the axis of symmetry are considered, as we do here.

On the other hand, in the resonant regime, with $|\tomegaE|\ll \mu \epsilon^2$, only neutrinos resonant with the unstable modes should be affected. If we include only the modes along the axis of symmetry, as we do here, we might expect only neutrinos along that direction to be affected. However, as discussed above, other modes resonate with neutrinos in these other directions, and therefore we expect again an effect on the entire angular distribution, but only when homogeneity is spontaneously broken along all directions, as generally expected. Therefore, a space-averaged distribution would be obtained only by breaking all symmetries, in marked difference to the fast instabilities of a single-crossed distribution, for which the most relevant unstable modes are usually directed along the axis of symmetry and are therefore captured even in a one-dimensional treatment. The more general topic of what is the space-time development of slow instabilities, in contrast to fast instabilities, as a consequence of their intrinsically resonant nature is discussed in detail in Part~II of this series of papers~\cite{Fiorillo:2025ank}.

To summarize, we have introduced a comprehensive treatment of the linear growth of slow instabilities in dense neutrino gases. We have thus brought the theory of slow flavor conversions to a comparable language and state as the theory of fast flavor conversions previously developed \cite{Fiorillo:2024bzm, Fiorillo:2024uki}, allowing one in principle to treat both on the same footing. In both cases, our framework allows for an intuitive understanding of what triggers the instability, namely the interaction between flavor waves and individual neutrino modes. Developing such an intuition may hopefully serve as a guide to tackle the much more complex problem of understanding what is the practical outcome of these instabilities in realistic astrophysical environments.

\acknowledgments

We thank Basudeb Dasgupta, Ian Padilla-Gay, Shashank Shalgar, G\"unter Sigl, Meng-Ru Wu, and Zewei Xiong for comments on the manuscript that have led to significant clarifications.
DFGF is supported by the Alexander von Humboldt Foundation (Germany),
whereas GGR acknowledges partial support by the German Research Foundation (DFG) through the Collaborative Research Centre ``Neutrinos and Dark Matter in Astro- and Particle Physics (NDM),'' Grant SFB-1258\,--\,283604770, and under Germany’s Excellence Strategy through the Cluster of Excellence ORIGINS EXC-2094-390783311.

\bibliographystyle{JHEP}
\bibliography{Biblio.bib}

\end{document}